\documentclass[prd,preprintnumbers,
eqsecnum,floatfix,letterpaper,superscriptaddress,nofootinbib]{revtex4}
\usepackage{amsfonts,amsmath,units,wasysym,epsfig,graphicx,verbatim,color,subfigure,graphicx}
\usepackage{amsmath}
\usepackage{latexsym}
\usepackage{amssymb}
\usepackage{amsfonts}
\usepackage{mathtools}
\usepackage{bm}
\usepackage{color}
\usepackage{float}
\usepackage{tikz}
\usepackage{adjustbox}
\usepackage{array}

\def\be{\begin{equation}}
\def\ee{\end{equation}}
\def\bea{\begin{eqnarray}}
\def\eea{\end{eqnarray}}

\def\no{\nonumber}
\def \D{{\cal D}}
\def \G{{\cal G}}
\def \SG{{\bf G}}
\def \VG{{\mathcal V}_{\G}}
\def \Vp{{\mathcal V}_\perp}
\def \S{{\cal S}}
\def \P{{\cal P}}
\def \N{{\cal N}}
\def \Msun {M_\odot}
\def \A{{\mathcal A}}
\def \Ab{{\bf A}}
\def \Ub{{\bf U}}
\def \Vb{{\bf V}}
\def \Sgm{{\bf \Sigma}}
\def \a{\alpha}
\def \b{\beta}
\def \M{{\mathcal M}}
\def \fl{f_{\rm lower}}
\def \fu{f_{\rm upper}}
\def \eps{\epsilon}
\def \x{{\bf x}}
\def \y{{\bf y}}
\def \h{{\bf h}}
\def \n{{\bf n}}

\def \s{{\bf s}}
\def \sp{{\s_\perp}}
\def \ts{{\tilde s}}
\def \prj{\mathfrak{p}}

\def \e{{\bf e}}
\def \v{{\bf v}}
\def \w{{\bf w}}

\begin{document}

\newcommand{\IUCAA}{Inter-University Centre for Astronomy and
  Astrophysics, Post Bag 4, Ganeshkhind, Pune 411 007, India}

\newcommand{\IITK}{Department of Physics, Indian Institute of Technology, Kanpur 208016, India}

\newcommand{\NIKHEF}{Nikhef - National Institute for Subatomic Physics, Science Park, 1098 XG Amsterdam, Netherlands}

\newcommand{\UVA}{Institute for High-Energy Physics, University of Amsterdam, Science Park, 1098 XG Amsterdam, Netherlands}

\newcommand{\WSU}{Department of Physics \& Astronomy, Washington State University, 1245 Webster, Pullman, WA 99164-2814, U.S.A.}

\newcommand{\IISERP}{Indian Institute of Science Education and Research Pune, Dr. Homi Bhabha Road, Pashan,  Pune  411008, India}

\newcommand{\AEI}{Max Planck Institute for Gravitational Physics (Albert Einstein Institute), Hannover, Germany} 

\newcommand{\sd}[1]{{\textcolor{red}{\bf{SD: #1}} }}

\newcommand{\sukanta}[1]{{\textcolor{magenta}{\bf{SB: #1}} }}

\newcommand{\prasanna}[1]{{\textcolor{blue}{\bf{Prasanna: #1}} }}

\newcolumntype{C}[1]{>{\centering\let\newline\\\arraybackslash\hspace{0pt}}m{#1}}

\author{Prasanna Joshi}
\affiliation{\IISERP}
\email{joshi.prasanna@students.iiserpune.ac.in}

\author{Rahul Dhurkunde}
\affiliation{\IISERP}
\affiliation{\AEI}
\email{rahul.dhurkunde@aei.mpg.de}

\author{Sanjeev Dhurandhar}
\affiliation{\IUCAA}
\email{sanjeev@iucaa.in}

\author{Sukanta Bose}
\affiliation{\IUCAA}
\affiliation{\WSU}
\email{sukanta@iucaa.in}

\date{\today}

\title{An optimal $\chi^2$ discriminator against modelled noise-transients in interferometric data in searches for binary black-hole mergers
}


\date{\today}

\begin{abstract}
A vitally important requirement for detecting gravitational wave (GW) signals from compact coalescing binaries (CBC) with high significance is the reduction of the false-alarm rate of the matched-filter statistic.
The data from GW detectors contain transient noise artifacts, or glitches, which adversely affect the 
performance of search algorithms, especially, for finding short-lived astrophysical signals, by producing false alarms, 
often with high signal-to-noise ratio (SNR).
These noise transients particularly affect the CBC searches, which are typically implemented by cross-correlating detector strain data with theoretically modeled waveform templates, chosen from a template-bank that is densely populated to cover the source parameter ranges of interest. Owing to their large amplitudes, many of the glitches can 
produce detectably large peaks in the SNR time-series -- termed as triggers --
in spite of their small overlap with the templates. Such glitches contribute to the false alarms. Historically, the traditional $\chi^2$ test has proved quite useful in distinguishing triggers arising from CBC signals and those caused by glitches.
\par

In a recent paper, a unified origin for a large class of $\chi^2$ discriminators was formulated, along with a procedure to construct an optimal $\chi^2$ discriminator, especially, when the glitches can be modeled. A large variety of glitches that often occur in GW detector data can be modeled as sine-Gaussians, with quality factor and central frequency, ($Q,f_0$), as parameters. An important feature of a sine-Gaussian glitch is that there is a lag between its time of occurrence in the GW data and the time of the trigger it produces in a templated search. Therefore, this time-lag is the third parameter used in characterizing the glitch. The total number of sampled points in the glitch parameter space is associated with the degrees of freedom (d.o.f.) of the $\chi^2$. We use Singular Value Decomposition to identify the most significant d.o.f.s, which helps in keeping the computational cost of our $\chi^2$ down. Finally, we utilize the above insights to construct a $\chi^2$ statistic that optimally discriminates between sine-Gaussian glitches and CBC signals. We also use Receiver-Operating-Characteristics to quantify the improvement in search sensitivity when it employs the optimal $\chi^2$ compared to the traditional $\chi^2$.
The improvement in detection probability is by a few to several percentage points, near a false-alarm probability of a few times $10^{-3}$, and holds for binary black holes (BBHs) with component masses from 
several to a hundred solar masses.
Moreover, the glitches that are best discriminated against are those that are like sine-Gaussians with $Q\in [25,50]$ and $f_0\in [40,80]$Hz.
\end{abstract}

\maketitle

\section{Introduction}
Great strides have been taken by modern technology in the past several decades which has allowed building of highly sensitive gravitational wave (GW) laser interferometric detectors. These are now capable of measuring GW strain sensitivities of $h \sim 10^{-22}$ or $10^{-23}$, where $h$ is the metric perturbation of the GW. The heroic
experimental efforts undertaken by physicists all over the world have
finally culminated with the first direct observation of a GW signal announced by the Laser Interferometer Gravitational Wave Observatory
(LIGO) project \citep{LIGO,GW150914}. On September 14, 2015, the two
LIGO interferometers at Hanford (Washington) and Livingston
(Louisiana), simultaneously measured and recorded strain data that
indicated the presence of a GW signal emitted by a coalescing binary
system containing two black-holes of masses of about  $36 M_\odot$ and $29 M_\odot$ at an average luminosity distance of $410$ Mpc. Since the announcement of the
first GW observation, more detections have been made by both LIGO and
the Virgo detectors, and it is expected that soon the KAGRA
interferometer in Japan \citep{KAGRA} will join the network in making astronomical observations.  We are now just beginning to explore the
observational capabilities offered by GWs, which promise to unveil
secrets of the Universe inaccessible by any other means
\citep{Thorne1987}. Future efforts are on to construct ever more sensitive GW detectors which will probe even deeper into the cosmos and complement the observations from electromagnetic astronomy thus giving us a more complete picture of the universe.
\par
 Detector data is neither Gaussian nor stationary. Non-Gaussianity and non-stationarity can arise from various components of the detector itself or the environment. Detection of GW signals crucially depends on comprehensively addressing the non-Gaussianity and non-stationarity of detector noise~\cite{Martynov2016} and the implementation of effective measures for discriminating noise artifacts from true signals (see, e.g., Ref.~\cite{Aasi2015}). In this work we focus on signals in ground-based detectors arising from compact binary coalescences (CBCs) involving black holes or neutron stars. These signals are transient, lasting between a fraction of a second to several minutes and can be adequately modelled with the help of post-Newtonian approximations and numerical relativity. While our primary focus here is on non-spinning BBHs, the basic ideas in this work can be extended to CBCs with spins and a wider distribution of masses. For signals that can be well modelled, matched-filtering is the commonly employed technique \citep{Helstrom} 
 -- a method that has been successfully
 applied to identify CBC signals buried in detector noise. Since the signals depend on several parameters a bank of templates densely covering the parameter space is employed ~\citep{SD91,DS94}. 
 However, just matched filtering by itself is not sufficient to identify a signal because the data contains non-Gaussianities and transient noise artifacts, also termed as glitches. 
 Even when the overlap of the glitches with the templates in the bank is small, the glitches themselves can be loud enough to produce triggers, which then run the risk of 
 being misinterpreted as signal-based. 
 In order to remedy this situation vetos or $\chi^2$ discriminators have been used. The traditional $\chi^2$ discriminator \citep{Allen2004} tends to distinguish  between a signal and a glitch by producing a high (low) value of the $\chi^2$ statistic if the trigger arises from a glitch (signal).
 The statistic is constructed based on the way the power in the frequency domain is distributed in various frequency bins by dividing the data into several frequency bins and checking whether this power distribution is consistent with that of the signal. Accordingly, a quantitative measure is defined - a $\chi^2$ statistic - based on the above considerations.
 
 However, this is not the only $\chi^2$ that is possible. It has been shown in \citep{DGGB2017} that a plethora - in fact an infinity - of such $\chi^2$ statistics can be constructed. The question addressed in \citep{DGGB2017} is what is a $\chi^2$ (in this context)? We briefly summarize its main results here. Consider the (function) space of all possible detector data trains $\D$ over an observation time $T$, with the scalar product defined by the power-spectral density (PSD) of the detector noise. $\D$ is a Hilbert space. A GW signal, a noise realisation, and a specific data train are all vectors in $\D$. So also is every template in a template-bank, with the additional property that 
 it has a unit norm. 
 A $\chi^2$ statistic amounts to assigning a relatively low-dimensional (say a few to 100) subspace $\S$ to each template vector in $\D$ such that the subspace $\S$ is orthogonal to that template vector. 
 Then the $\chi^2$ associated with any data vector in $\D$, and a given template, is just the norm-squared of the projection of that vector onto the subspace $\S$ assigned to that template. 
 Furthermore, the number of degrees of freedom of the $\chi^2$ is just the dimension of $\S$. For a fixed dimension of $\S$, each $\chi^2$ statistic amounts to constructing a vector bundle over 
 the signal manifold or the parameter space $\P$. The traditional $\chi^2$ is just one choice of the subspaces $\S$ resulting in one such vector bundle. 
 Since $\S$ can be chosen in a plethora of ways, a large number of such $\chi^2$ are possible. We have then a large freedom in our choice of discriminatory tests and this freedom can be utilised in a fruitful way to optimise the signal search statistic. This can be certainly done for glitches that can be modelled. 
 
 As remarked earlier, the detector data is glitchy. The glitches may be  classified based on their morphology in a time-frequency map. 
 A family of glitches that frequently occur in the detector data, and are troublesome in some ways, have the structure of sine-Gaussians - 
 or those that can be modelled as sine-Gaussians. 
 In this paper we focus on such glitches. Our aim in this paper is to design a $\chi^2$ statistic that is optimal for these type of glitches. 

The question to be addressed here is  how to make the $\chi^2$ optimal?
It is clear that we will get a high value of $\chi^2$ if we align the subspace $\S$ along the glitches, so that the glitches have maximum projection on $\S$. (We must also satisfy the requirement that $\S$ must be  orthogonal as well to the template, but this is easily achieved because $\dim (\S) \ll \dim (\D)$ - there is enough ``room" to orient $\S$.)
However, a reasonable sampling of the glitches -- giving at least a projection of, say, 90$\%$ -- results in a large number of glitch vectors: We find this number to be a few thousand, typically. This will make $\dim (\S) \sim$ few thousand, which is the number of degrees of freedom for the $\chi^2$. This would push up the computational cost. 
Our strategy is then to approximate the subspace spanned by the glitch vectors by a lower dimensional subspace of say less than 100. 
This is what we will choose as $\S$. We must then find the best approximation to the subspace spanned by the glitch vectors. 
This is achieved via the Eckart-Young-Mirsky theorem~\citep{EY1936}. It uses the Singular Value Decomposition (SVD) \citep{NumericalRecipes} \citep{Matrix_comp-Frob} to find the best approximation to a subspace of dimension $n$ with a subspace of dimension $m$, where $m < n$. The details follow in later sections. There are several non-trivial steps involved - ensuring that $\S$ is orthogonal to the trigger template, dealing with a general scalar product because of the coloured PSD, etc. 
We describe these aspects in section \ref{construction}.
\par

We accordingly construct an optimal $\chi^2$ to disciminate against sine-Gaussian glitches. We call it the {\it optimal  sine-Gaussian} $\chi^2$ and denote it by $\chi^2_{\rm SG}$. 
We perform simulations of CBC signals, sine-Gaussian glitches, and detector noise and use them to construct Receiver-Operating-Characteristics for quantifying the improvement in search sensitivity when it employs the optimal sine-Gaussian $\chi^2$ compared to the traditional $\chi^2$.
As we show below, the improvement in detection probability is by a few to several percentage points, near a false-alarm probability of a few times $10^{-3}$, and holds for binary black holes  with component masses from 
several to a hundred solar masses.
Moreover, the glitches that are best discriminated against are those that are like sine-Gaussians with $Q\in [25,50]$ and $f_0\in [40,80]$Hz.
\par

The paper is organised as follows. In section \ref{geom} we describe earlier work pertinent to the problem we discuss here; we give a brief review of matched filtering, the unified $\chi^2$ and sine-Gaussians. In section \ref{construction} we describe in detail the steps required to construct an optimal $\chi^2$ that will discriminate against sine-Gaussian glitches. This involves sampling the parameter space of sine-Gaussians with sufficient number of points, whittling down this number with the help of the SVD algorithm in order to obtain the best low-dimensional approximation to the vector space spanned by the sampled sine-Gaussians (Eckart-Young-Mirsky theorem), adaptation of the SVD to coloured noise, etc. In section \ref{sec:results} we
apply the aforementioned construction to compute the optimal sine-Gaussian $\chi^2$ on simulated CBC signals and sine-Gaussian glitches. We compare the performance of detection statistics employing the new $\chi^2$
and the traditional
$\chi^2$
on the same simulations. These comparisons are described 
with the help of $\chi^2$ versus SNR plots and  Receiver Operating Characterstics (ROC) curves. In the final section \ref{concl} we conclude with a discussion on future applications.

\section{The underlying geometrical structure}
\label{geom}
\subsection{The matched filtering programme}
\label{MF}

Consider two data trains (or functions), $x(t)$ and $y(t)$, defined over a time interval $[0, T]$ of duration $T$. The data trains form a vector space $\D$. As vectors in $\D$, they will be denoted in boldface -- $\x$ and $\y$. Let $n(t)$ be the noise in the detector, which is a stochastic process defined over the data segment, has ensemble mean of zero, and is stationary in the wide sense. A specific noise realisation is a vector $\n \in \D$ - $\n$ is in fact a random vector. Its PSD is denoted by $S_h (f)$. The scalar product of $\x$ and $\y$, is written conveniently in the Fourier domain. If $\tilde{x}(f)$ and $\tilde{y}(f)$ are the Fourier representations of $\x$ and $\y$, then the scalar product is given by:
\be 
 (\x ,\y) = 4 \Re \int_{\fl}^{\fu}~ df \frac{\tilde{x}^* (f) \tilde{y}(f)}{ S_{h}(f)} \,,
\label{scalar} 
\ee
where integration is carried out over the band-width $[\fl, \fu]$. This construction makes the space of data segments a Hilbert space - a $L_2$ space with measure $d \mu \equiv df / S_h (f)$. We denote this space by $\D = L_2([0, T], \mu)$.  

The most commonly used post-Newtonian (PN) approximant is TaylorF2, which is computed in the Fourier domain using the stationary phase approximation. We choose this approximant for the signal in this work, which can be straightforwardly generalized to other waveform models. The general form of the signal, denoted by $h$, is
\be
{\tilde h} (f) = \A ~f^{-7/6}~e^{- i \psi(f)} \,,
\ee
where the overall amplitude $\A$ depends on the binary component masses, the source distance, sky position and the  orientation of the binary orbit relative to the detector. The phase $\psi (f)$ is computed to 3.5PN order explicitly~\cite{Buonanno2009}, and depends on the coalescence time and phase, $t_c, \phi_c$,  respectively, and the mass parameters. We will view these waveforms as vectors in $\D$ and denote them by the boldfaced letter $\h$.

The Newtonian waveform, which is  simple, even if somewhat inaccurate, is nevertheless useful for illustrating the key ideas in this work. The normalized Newtonian inspiral binary waveform in the Fourier domain is given by: 
\begin{align}
{\tilde h}(f; t_c, \tau_0, \phi_c)= \N f^{-\frac{7}{6}} e^{-i \psi_N (f; t_c, \tau_0, \phi_c)} \,,
\end{align}
where $\N$ is a normalization constant determined by setting $(\h, \h) = 1$.  The phase $\psi_N (f)$ is given by:
\begin{align}
\psi_N (f; t_c, \tau_0, \phi_c)= 2 \pi f t_{c}  + \frac{6 \pi f_s \tau_0}{5} \left( \frac{f}{f_s} \right)^{-5/3} - 
\phi_c - \frac{\pi}{4}\,.
\end{align}
 Furthermore, we have expressed the phase in terms of a parameter more suited to this work than the chirp mass~\cite{SD91}, namely, the chirp time $\tau_0$~\cite{SD91,DS94}.  Physically, $\tau_0$ is approximately the time taken for the binary to coalesce starting from some fiducial frequency $f_{a}$. We take this fiducial frequency to be near about the lower end of the range of central frequencies $f_0$ of the sine-Gaussians that we will consider. Taking $f_0 = 30$ Hz we obtain,
\be
\tau_0 =\frac{5}{256 \pi f_0} (\pi \M f_0)^{-5/3} \simeq 5.085 \left( \frac{f_0}{30 ~{\rm Hz}} \right)^{-8/3} \left( \frac{\M}{5 \Msun} \right)^{-5/3} {\rm sec} \,,
\label{chrpt}
\ee
where $\M = \mu^{3/5} M^{2/5}$ is the chirp mass, $\mu$ and $M$ being the reduced and the total mass, respectively. Also, $\Msun$ denotes the mass of the Sun. We have set $G = c = 1$. 
\par
The signal $\s$ in the data is just an amplitude $A$ multiplying the normalized waveform $\h$; thus, $\s = A \h$. The data vector, which we denote by $\x$, is then $\x = \s + \n$, when a signal is present; In the absence of a signal it is just noise, i.e., $\x = \n$. The match $c$ (correlation) is the scalar product between the data $\x$ and a (normalized) template  $\h$, that is, $c = (\x, \h)$, which is then a function of the template parameters. In the analysis of the data for searching signals the match is maximized over template parameters and compared with a preset threshold. In practice, for the parameters $t_c, \phi_c$, the templates need to be only defined at $\phi_c = 0$ and $\phi_c = \pi/2$, and for $t_c = 0$. This is because the search over these parameters can be done efficiently using quadratures for $\phi_c$ and the FFT algorithm for $t_c$ in a continuous fashion. The search over the mass parameters is carried out with a densely sampled discrete bank of templates so that chance of missing out a signal is low. 

\subsection{The unified $\chi^2$}

The $\chi^2$ discriminator is defined so that its value at the signal is zero and for Gaussian noise it has a $\chi^2$ distribution with a certain number of degrees of freedom. The $\chi^2$  test for the trigger template $\h$ is defined by choosing a finite dimensional subspace $\S$ of dimension $p$ such that for any $\v \in \S$, we must have $(\v, \h) = 0$, that is, $\S$ as a subspace is orthogonal to $\h$. Then the $\chi^2$ pertaining to the template $\h$ is just the square of the $L_2$ norm of the data vector $\x$ projected onto $\S$. Specifically, we decompose the data vector $\x \in \D$ as,
\be
\x = \x_{\S} + \x_{\S^\perp} \,,
\ee
where $\S^{\perp}$ is the orthogonal complement of $\S$ in $\D$. $\x_{\S}$ and $\x_{\S^\perp}$ are projections of $\x$ into the subspaces $\S$ and $\S^{\perp}$, respectively. We may write $\D$ as a direct sum of $\S$ and $\S^{\perp}$, that is, $\D = \S \oplus \S^{\perp}$.
\par
Then the statistic $\chi^2$ is,
\be
\chi^{2} (\x) = \| \x_{\S} \|^2 \,.
\ee
Given any orthonormal basis in $\S$ say $\e_{\a},~~\a = 1, 2, ..., p$ so that $(\e_{\a}, \e_{\b}) = \delta_{\a \b}$, where $\delta_{\a \b}$ is the Kronecker delta, we easily verify its properties:

\begin{enumerate}

\item For a general data vector $\x \in \D$, we have:
\be
\chi^2 (\x) = \| \x_{\S} \|^2 = \sum_{\a = 1}^p |(\x, \e_{\a})|^2 \,, 
\ee   

\item Clearly, $\chi^2 (\h) = 0$, because the projection of $\h$ into the subspace $\S$ is zero or $\h_{\S} = 0$. 

\item Now let us take the noise $\n$ to be stationary and Gaussian with PSD $S_h(f)$ and mean zero. Then the following is valid:
\be
\chi^2 (\n) = \| \n_{\S} \|^2 = \sum_{\a = 1}^p |(\n, \e_{\a})|^2 \,.
\ee
Observe that the random variables $(\n, \e_{\a})$ are independent and Gaussian, with mean zero and variance unity. This is because $\langle (\e_{\a}, \n) (\n, \e_{\b}) \rangle = (\e_{\a}, \e_{\b}) = \delta_{\a \b}$, where the angular brackets denote ensemble average (see \citep{Creighton:2011zz} for proof). Thus, $\chi^2 (\n) $ has a $\chi^2$ distribution with $p$ degrees of freedom. 
\end{enumerate}

For the ease of  calculations, one is free to choose any {\em orthonormal} basis of $\S$. In an orthonormal basis  the statistic is manifestly $\chi^2$ since it can be written as a sum of squares of independent Gaussian random variables, with mean zero and variance unity.

However, in the context of CBC searches, we are in a more complex situation. We do not have just one waveform but a family of waveforms that depend on several parameters, such as masses, spins and other kinematical parameters. We denote these parameters by $\lambda^a, ~~ a = 1, 2, ..., m$. As before, we may assume the waveforms to be normalized, i.e., $\| \h (\lambda^a) \| = 1$. (We have excluded the amplitude $\A$, but it can be easily reinstated. This is in fact the manifold traced out by the templates and is a sub-manifold of the unit hypersphere in $\D$.) Then the waveforms trace out an $m$-dimensional manifold $\P$ - the signal manifold -  which is a submanifold of $\D$. We now associate a $p$-dimensional subspace $\S$ orthogonal to the waveform $\h (\lambda^a)$ at each point of $\P$ - we have a $p$-dimensional vector-space ``attached" to each point of $\P$. When done in a smooth manner, this construction produces a fibre bundle with a $p$-dimensional vector space attached to each point of the $m$ dimensional manifold $\P$. The fibre bundle so obtained is a vector bundle of dimension $m + p$. We have, therefore, found a very general mathematical structure for the $\chi^2$ discriminator. Any given $\chi^2$ discriminator for a signal waveform $\h (\lambda^a)$ is the $L_2$ norm of a given data vector $\x$ projected onto the fibre $\S$ at $\h (\lambda^a)$. 
\par
It can be easily shown that the traditional $\chi^2$ falls under the class of unified $\chi^2$. This is done by exhibiting the subspaces $\S$ or by exhibiting basis field for $\S$ over $\P$; the conditions mentioned above must be satisfied by $\S$. In \citep{DGGB2017} such a basis field has been  given explicitly.

\subsection{Sine-Gaussian glitches}

\begin{figure}[H]
\centering
\includegraphics[width = 3in]{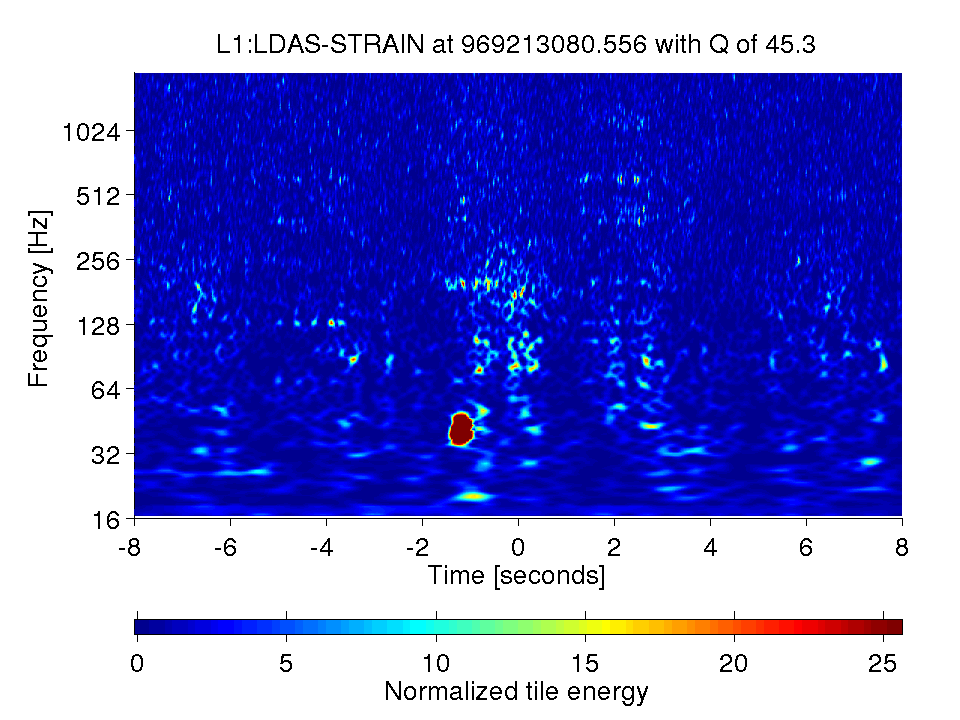}
\includegraphics[width = 3in]{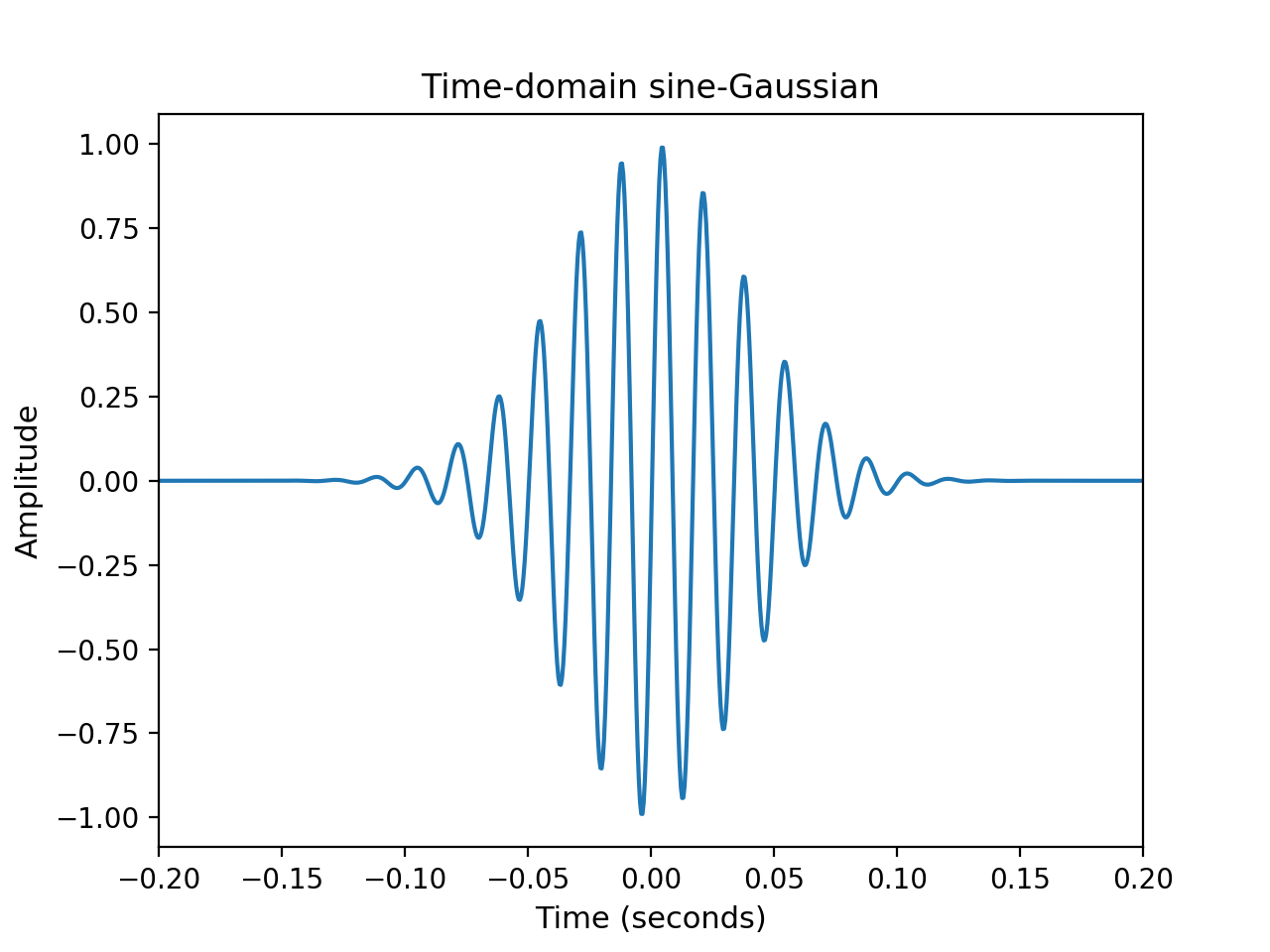}
\caption{The figure on the left shows a sine-Gaussian glitch in real data. The spectrogram shows a noise transient, located at
  approximately -1.2 sec in the time coordinate used above, that 
appeared in the gravitational-wave channel of the LIGO
  detector in Livingston (L1) during its sixth science run (S6). This
  figure was made by Omega scan and shows that the loudest sine-Gaussian
 component of this transient has $f_0 = 39.3$ Hz and $Q=45.3$. The figure on the right shows the modelled glitch with $f_0 = 60$ Hz, $Q = 20$ and $t_0 = 0$.
 }
\label{glitches}
\end{figure}

 Many transient bursts are represented suitably in the form of sinusoids with a Gaussian envelope \cite{Sourav}. We can model these glitches by using a sine-Gaussian model with central frequency $f_0$, central time $t_0$ and a  quality factor $Q$. 
 A glitch occurring in real data is shown below on the left panel in Fig.~\ref{glitches}; the panel on the right shows a modelled glitch.

The time-domain expression for a sine-Gaussian (shown in Fig.~\ref{glitches}) with central frequency $f_0$, quality factor $Q$ and central time $t_0$ is given by:
\begin{align}
     s(t) = s_0 e^{-(t-t_0)^2/\tau^2}\sin{2\pi f_0 (t - t_0)}\,,
\label{eq:sgt} 
\end{align}
where $s_0$ is the amplitude and $\tau$ is the decay time-constant related to the quality factor as $Q=2\pi f_0 \tau$. The frequency-domain expression can be obtained by Fourier transforming $s(t)$, and can be shown to be a Gaussian centered at $f_0$:
\begin{align}
    \tilde{s}(f) = \kappa ~ e^{-\frac{(f-f_0)^2Q^2}{4f_0^2}}.
\label{FT_sg}    
\end{align}
where $\kappa$ is a normalisation constant. If we demand that
\begin{align}
  4 \int_0^{\infty} df~ |\ts (f)|^2  = 1 \,,  
\end{align}
then $\kappa =(Q /2 f_0)^{1/2} (1 /2 \pi)^{1/4}$. 
Here we have set the central time $t_0$ of the sine-Gaussian to be zero. However, for a non-zero $t_0$ the $\tilde{s} (f)$ in (\ref{FT_sg}) will be merely multiplied by the factor $e^{-2\pi i f t_0}$.

One can conceptualize the family of glitches, say $\G$, as a manifold. In fact, it is a three-dimensional manifold with coordinates $(t_0, f_0, Q)$. Indeed, it can even be equipped with a metric, which is a map from coordinate differences of neighboring unit-norm sine-Gaussians to the fractional drop in their match~\cite{Balasubramanian:1995bm,Owen:1995tm}.
It can be described by the line-element on that manifold,
\begin{align}
ds^2 = 4\pi f_0^2 \left (1 + \frac{1}{Q^2} \right)dt_0^2 + \dfrac{2+Q^2}{4f_0^2}df_0^2 + \dfrac{1}{2Q^2}dQ^2 - \dfrac{1}{f_0 Q}df_0 dQ\,.
\label{metric_1}
\end{align}
(Note that $ds$ does not describe an infinitesimal change in $s$ of Eq.~(\ref{eq:sgt})!)
There is a cross term in the metric in these coordinates. A set of parameters that we find useful is $f_0 \rightarrow \omega_0 = 2 \pi f_0$ and $\nu = 1/\tau$. Then $Q \rightarrow \omega_0/\nu$. In these new coordinates we obtain the metric in a diagonal form as:
\begin{align}
	ds^2 = (\nu^2 + \omega_0^2) dt_0^2 + \frac{1}{4\nu^2} d\omega_0^2 + \frac{1}{2\nu^2} d\nu^2 \,.
\label{metric_2}
\end{align}
We will make use of these metric forms for uniformly sampling the space $\G$ of sine-Gaussians so that they have adequate projection on the subspaces $\S$.

Two comments are in order. First, this metric is a little different from the one in \citep{Sourav}. The metric here is derived by taking the real part of an integral, as in Eq. (\ref{scalar}); 
whereas the one in  Ref.~\citep{Sourav} is derived from the modulus of that integral. Accordingly, we have an extra $\omega_0^2$ term multiplying $dt_0^2$ -- otherwise the metrics are identical. The two metrics serve different purposes in their application. Second, 
$\G$ is not a submanifold of $\D$ in the strict sense because the metrics (\ref{metric_1}) and (\ref{metric_2}) are not induced from the metric on $\D$. The metric on $\D$ derived from the scalar product Eq. (\ref{scalar}) depends on the PSD $S_h (f)$. However, if $\D$ had an Euclidean metric (or if the noise was white), then the metric on $\G$ would be the induced metric and $\G$ would be a sub-manifold of $\D$. However, since ultimately, we only require the sampling to be approximately uniform, these metrics work for us. 

\section{Optimising the $\chi^2$ for sine-Gaussian glitches}
\label{construction}

In this section we describe how to construct the subspace $\S$ that is optimal for discriminating against sine-Gaussian glitches associated with a specific trigger template $\h$. The method operationally uses the Singular Value Decomposition (SVD) algorithm in order to arrive at $\S$. There are essentially three steps involved:
 
\begin{enumerate}
 \item Sample the parameter space $\G$ of sine-Gaussians so that any specific sine-Gaussian not in the sample has adequate projection on the vector space spanned by the sampled vectors. We call this space $\VG$ which is a subspace of $\D$. When a reasonably high projection is desired, $\G$ must be sampled densely. We will also endeavour to do it uniformly for the sake of economy. 
 \item Piece together a matrix consisting of the sampled sine-Gaussian row vectors. These row vectors need to be appropriately modified so that one gets the desired $\S$. There are several steps here which will be described in the text that follows.
 \item Applying SVD to the space spanned by the appropriate row vectors will obtain for us the best possible approximation of lower dimension. This will be our subspace $\S$. Since the scalar product on $\D$ is not strictly in the Euclidean form (in Fourier space it is scaled by the inverse of the PSD), appropriate modifications must be made to the input matrix and also to the output matrix so that the SVD only ``sees" an Euclidean scalar product. Further the output matrix containing right singular vectors needs to be unwhitened so that the resulting vectors span $\S$ - in fact they form an orthonormal basis of $\S$. We are actually in the realm of the generalised SVD. 
\end{enumerate}

 We now elaborate on these steps in the subsections that follow.
 
\subsection{Sampling the space of sine-Gaussians}
\label{sampl}

It is observed that, when a CBC template is triggered by a sine-Gaussian glitch, the trigger occurs with a time-lag $t_d$ after the glitch~\cite{Tito,Bose2016,2016JPhCS.716a2007B}. Depending on how low $f_0$ is, his time-lag can be as large as the length of the chirp waveform. 
For aLIGO, if $f_0$ is low, say, a few tens of Hz, the time-lag will be of the order of several minutes. This is because the sine-Gaussian glitch is essentially narrow band and matches with the template in the neighbourhood of the frequency $f_0$. If $f_0$ is low, then the chirp template takes significant time to reach coalescence -- which is in fact the time-lag. In-depth analysis has been performed on this issue:
As shown in Ref.~\cite{Bose2016}, the time-lag $t_d$ is approximately given by
\begin{align}
	t_d \simeq \tau_0\left(1 - \frac{16}{3Q^2} \left(\zeta + \frac{2}{3}\right)\right)\,,
\label{tlag} 
\end{align}
where $\tau_0$ is the chirp time given by Eq. (\ref{chrpt}) and $\zeta$ is the logarithmic derivative of the noise PSD $S_h (f)$ evaluated at $f_0$.  Since we have taken $Q > 5$, the term involving  $1/Q^2$ is very small and may be ignored  compared to unity. Therefore we may write,
\begin{align}
    t_d \simeq \tau_0 = \frac{5}{256\pi f_0} (\pi \mathcal{M} f_0)^{-5/3}\,.
    \label{td_approx}
\end{align}
Here the Newtonian approximation to the waveform has been used to compute $t_d$. This is justified well below. 

Now if the glitch occurs at $t = 0$, the trigger will occur at time $t_d$. Or, viewing the situation the other way, if the trigger occurs at $t = 0$ for a given template in the bank, the glitch must be at $t = - t_d$, which is a function of $f_0$ and $Q$ (and, of course, the template masses, mainly in the combination $\mathcal{M}$). But since we do not know {\it a priori} the parameters of the glitch, our strategy is to sample  
those sine-Gaussians that would give rise to a trigger at $t = 0$. 
Thus, we only need to sample the 2-dimensional surface  $t_0 = - t_d (f_0, Q)$ instead of the larger 3-dimensional manifold $\G$. This is easily done by computing the induced metric on this surface by substituting the expression for the surface into the metric given in Eq. (\ref{metric_2}). 
\par

In our simulations that follow, we will employ the IMRPhenomP waveform approximant~\cite{Hannam:2013oca}.
(Although we limit the simulated BBHs to the non-spinning variety here, we plan to extend it to spinning BBHs in the future.)
Due to post-Newtonian corrections and other effects, the time-lag $t_d$ computed with the IMRPhenomP waveform will differ from the Newtonian one given in Eq.~(\ref{td_approx}) by a small amount, say, $\Delta t_0$. The Newtonian chirp time $\tau_0$ is the primary contributor to $t_d$. Therefore, $\Delta t_0$ will be small compared to $\tau_0$. Thus, geometrically speaking, we will be stepping out of the Newtonian surface $t_0 = - t_d (f_0, Q)$.  However, since we are sampling the full Newtonian surface, one may look for any sine-Gaussian in the surface close to the sine-Gaussian at $t_d + \Delta t_0$, provided it exists. It turns out that for the parameters considered here, the surface is such that the $t_0$ axis is almost parallel to this surface. This means that if we consider the sine-Gaussian {\it in the surface} with time-lag $t_d + \Delta t_0$, it is very close to the one outside the surface, albeit with a slightly different $f_0$, say, $f_0 + \Delta f_0$. From the metric in Eq.~(\ref{metric_2}) we see that the distance between these two sine-Gaussians is $\Delta s \simeq \Delta \omega_0 /2 \nu$, which is very small for the parameters studied. We have  numerically checked and found that $\Delta t_0 \lesssim 10$ milliseconds and the projection is better than 99 $\%$. This shows that our analysis is robust to small errors in $t_d$.

Based on the detector data, as well as convenience, we choose the following ranges for the parameters: $40~ {\rm Hz} \leq f_0 \leq 120$ Hz and $5 \leq Q \leq 50$. For these chosen ranges of parameters further simplifications of the metric are possible and they facilitate the sampling. First of all in Eq. (\ref{metric_2}) we can drop $\nu^2$ compared to $\omega_0^2$ in coefficient of $dt_0^2$. Also writing $z = (\omega_0 \M)^{-5/3}$, we get
 \bea
 ds^2 &=& \omega_0^2 dt_0^2 + \frac{1}{4\nu^2} d\omega_0^2 + \frac{1}{2\nu^2} d\nu^2  \, \no \\
 &\simeq& 2^{-14/3} dz^2 + \frac{1}{4\nu^2} d\omega_0^2 + \frac{1}{2\nu^2} d \nu^2 \,.
 \label{metric_3} 
 \eea
Further, we also find that the second term in the above metric can be written as
\be
\frac{d \omega_0^2}{4 \nu^2} = \frac{9 Q^2}{100} \frac{dz^2}{z^2} \,.
\ee
For templates with $\M \sim 10 M_{\odot}$ and for the values of $f_0$ and $Q$ considered, $z \sim 10^3$ or $10^4$, so the contribution of this term to the coefficient of $dz^2$ is $\sim 10^{-4}$ while the first term is $2^{-14/3} \sim 0.04$. Thus we may neglect the $d\omega_0$ term from the metric. Further, writing $y = \ln(\nu)$, we obtain
\begin{align}
	ds^2 = 2^{-14/3} dz^2 + \frac{1}{2} dy^2 \,.
\label{metric_4}
\end{align}
 We have finally arrived at a metric that is flat (i.e., the metric coefficients are independent of the coordinates). In this form of the metric our task becomes that much easier. 
 
Instead of setting up a rectangular lattice of points, it is more convenient to 
put up points along curves  $Q = $const. The equation for the curve with constant $Q$, in $y-z$ coordinates, can be derived from the relation:
\begin{align}
y &= -\frac{3}{5}\ln z - \ln Q - \ln \mathcal{M} \,.
\label{yz}
\end{align}
Then the second axis of the lattice is given by $z =$ const. We choose the grid in this manner because the boundaries of the region of the parameter space are inconvenient curves in $y-z$ coordinates. The grid points satisfy the following criteria:
\begin{enumerate}
\item The distance between the points is so adjusted that any sine-Gaussian in the parameter space has at least projection $\prj$ on some grid vector. We generally choose  $\prj \geq 0.8$ or $80 \%$. The projection $\prj$ translates to the mismatch $\eps = \sqrt{2(1 - \prj)}$. The choice of $\prj$ and the corresponding $\eps$ is summarized in Table \ref{tab:choices_of_p}.
\item  The grid points satisfy the condition that the distance between two adjacent points is the same, namely, $\sqrt{2} \eps$. This distance has been so chosen that the criterion 1 is satisfied. The metric given in Eq. (\ref{metric_4}) is used to accomplish this. The grid however is inclined.
\item The distance between grid points is chosen large enough so that there are minimum number of points in the grid while at the same time ensuring that criterion 1 is satisfied.
\end{enumerate}
\begin{table}
    \centering
    \begin{tabular}{|C{2cm}|C{2cm}|C{2cm}|C{2cm}|}
        \hline
        $\bf M_{min} (M_{\odot})$ & $\bf M_{max} (M_{\odot})$ & $\prj$ & $\eps$ \\
        \hline
        $10$ & $70$ & $0.80$ & $0.632$ \\
        \hline
        $70$ & $90$ & $0.85$ & $0.548$\\
        \hline
        $90$ & $100$ & $0.90$ & $0.447$\\
        \hline
        $100$ & $120$ & $0.95$ & $0.316$\\
        \hline
        $120$ & $130$ & $0.975$ & $0.224$\\
        \hline
        $130$ & $160$ & $0.99$ & $0.141$\\
        \hline
    \end{tabular}
    \caption{The above table contains the choice of value of $\prj$ and the corresponding value of $\eps$ for templates with total mass lying in the corresponding range.}
    \label{tab:choices_of_p}
\end{table}

\begin{figure}[H]
\centering
\includegraphics[width = 3.5in]{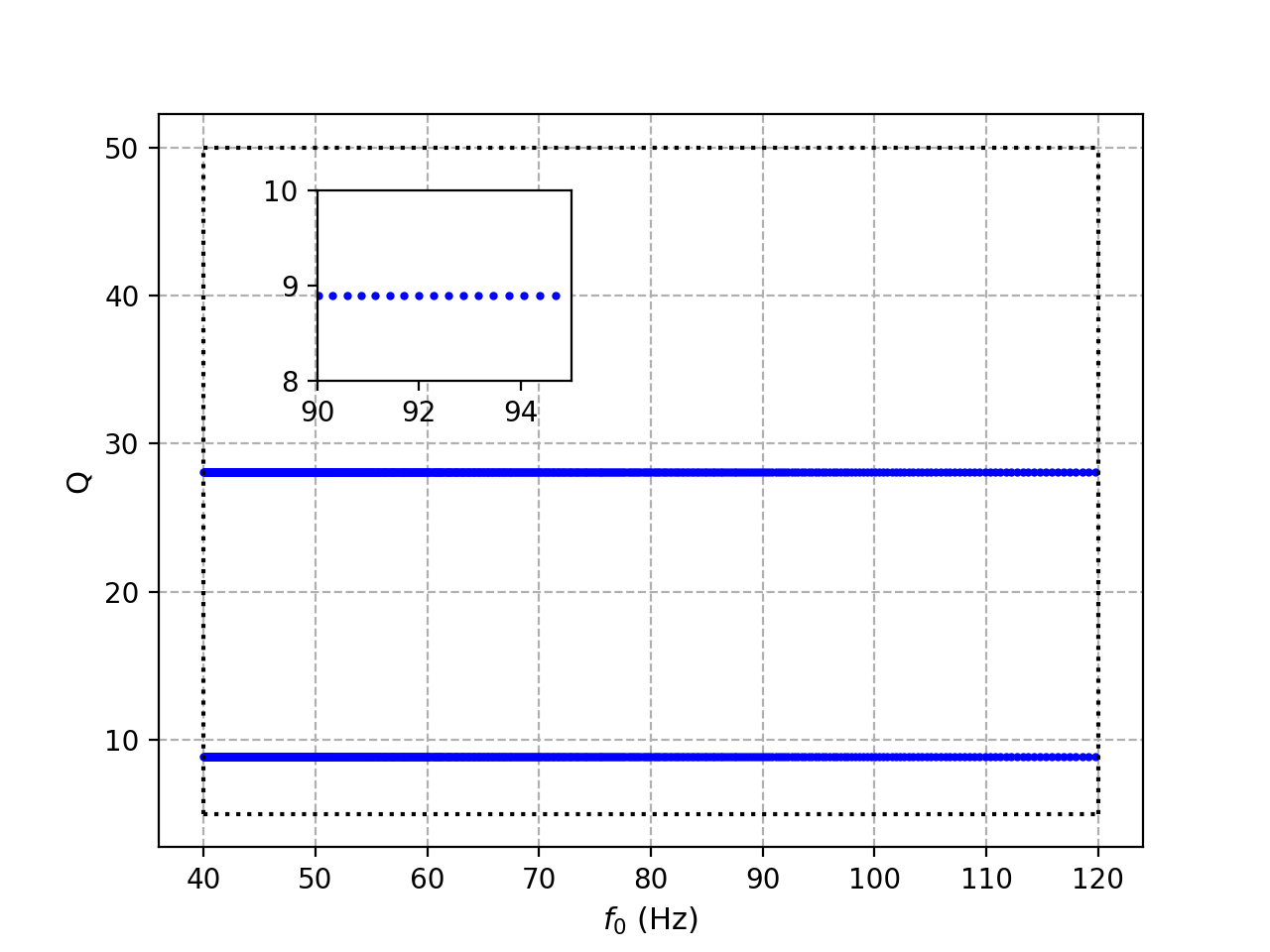}
\includegraphics[width = 3.5in]{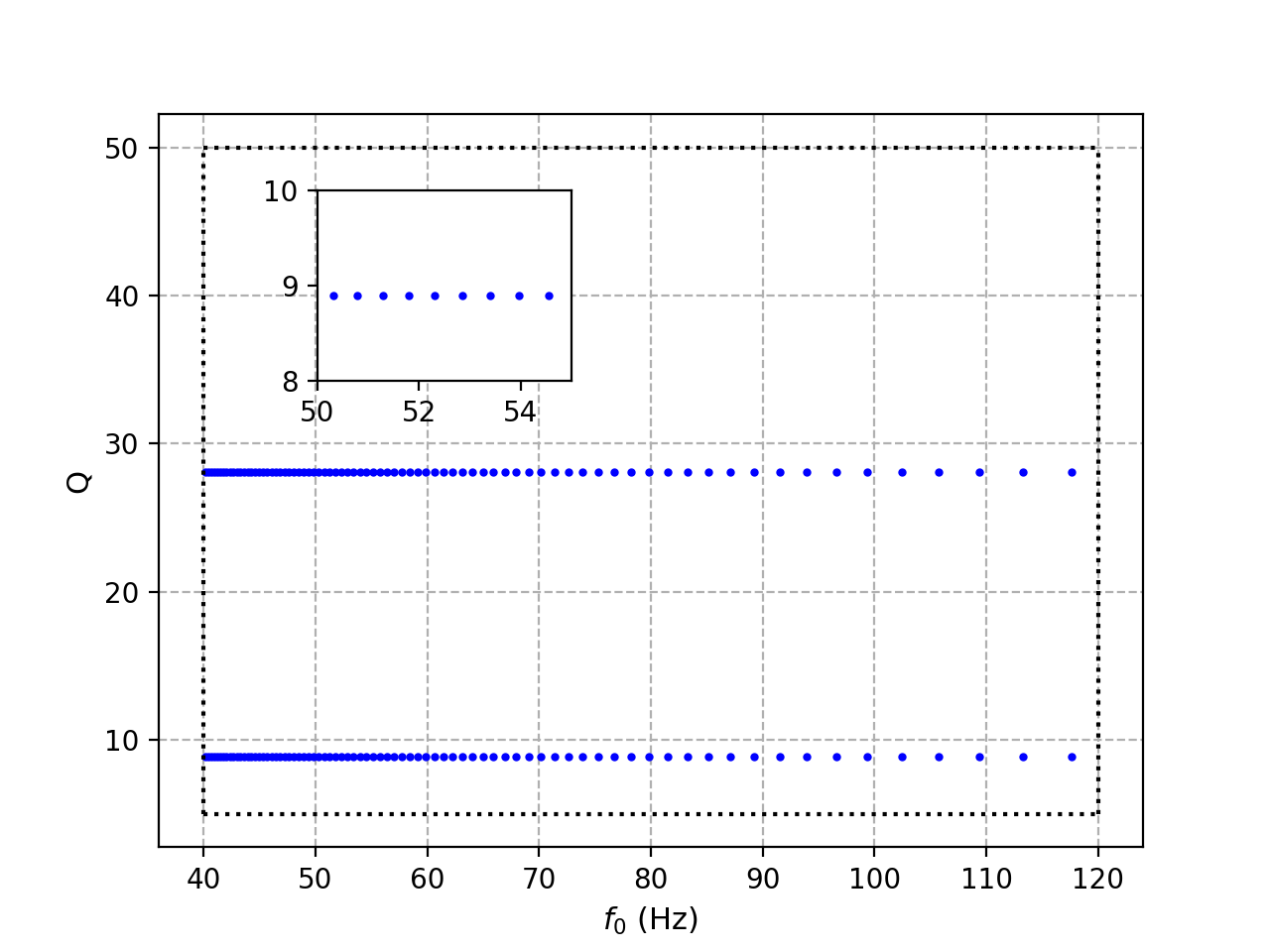}
    \caption{The figures shows uniformly sampled points in the parameter space $(f_0, Q)$ in the range $40 \leq f_0 \leq 120$ Hz, $5 \leq Q \leq 50$.  The minimum projection is 80\%. The figure on the left is for component masses of $7 M_{\odot}$ each and the total number of sampled points is 1288 (see the inset figures to note how closely spaced the neighboring points are). The figure on the right is for component masses of $25 M_{\odot}$ each and the number of points sampled is 156. 
    }
\label{smpl}    
\end{figure}

In $y-z$ coordinates the grid points are given by
\begin{align}
	y_{ij} = -\frac{3}{5}\ln z_j - \ln Q_i - \ln \mathcal{M}\,.
\end{align}
The distance between adjacent grid points is $\sqrt{2} \eps$. 
In Fig. \ref{smpl} we have shown the grid points in the $f_0-Q$ plane, $40 \leq f_0 \leq 120$ Hz, $5 \leq Q \leq 50$.  
The minimum projection is 80\%. The figure on the left is for individual masses of $7 M_{\odot}$ with the number of grid points being 1288. The figure on the right is for individual masses of $25 M_{\odot}$ with the number of grid points being 156. All these numbers are related to the area of the parameter space.
\par

We now compute the area $\A$ of the parameter space. The area element of the parameter space is found easily from the metric form Eq.~(\ref{metric_4}) and we set the limits on $y$ from Eq.~(\ref{yz}). The result is:
\bea
\A &=& 2^{-17/6} \int_{y_{\min}}^{y_{\max}} dy \int_{z_{\min}}^{z_{\max}} dz \, \no \\
&=& 2^{-17/6} (z_{\max} - z_{\min}) \ln \left( \frac{Q_{\max}}{Q_{\min}} \right) \,.
\eea
 Since $z$ scales as $\M^{-5/3}$ so does the area $\A$. Clearly, the number of grid points is proportional to the area of the parameter space. The area of the parameter space is $937.79$ for $7 M_{\odot}$ and $112.38$ for $25 M_{\odot}$.  


We remark that this is not the optimal way to sample the parameter space for a given projection $\prj$ - we could have obtained a smaller number of grid points by strictly choosing a square lattice or even a hexagonal lattice -- here there are about $10 \%$ more points than what we would have had for the square lattice of side $\sqrt{2} \eps$ (there is also a slight excess from boundary effects).  However, our basic goal here was to sample the parameter space adequately and we have done this in a convenient manner. In the text that follows, we use the SVD algorithm \citep{NumericalRecipes} \citep{Matrix_comp-Frob} to arrive at the best low-dimensional approximation to the subspace spanned by the sampled vectors. The SVD is expected to whittle down the subspace to appropriate number of dimensions and, thus, nullify the effects of oversampling.

\subsection{Preparing the input matrix for the SVD}
\label{inmatrix}

The sampled sine-Gaussians of section \ref{sampl} cannot be directly used in the present form in the SVD algorithm. 
\par
This is because:
\begin{itemize} 
\item The sine-Gaussians have central time $t_0 = 0$ and they need to be appropriately time adjusted with respect to the time of occurrence of the trigger. 
We will always take the trigger to occur at $t = 0$, and so the glitch must have occurred at time $- t_d$. Note that the $t_d$ depends on $f_0$ and $Q$. 
\item  We need to find the components of the sine-Gaussians orthogonal to the trigger template. This is achieved by subtracting out from each sine-Gaussian its component that is parallel to that template. The orthogonal components of the sine-Gaussians so resulting need to be further time-shifted appropriately by an amount $-t_d$. Finally, after these operations, the resulting vectors span a subspace of $\D$ that we will denote by $\Vp$ (we drop $\G$ to avoid clutter). The subspace $\S$ will turn out to be a subspace of $\Vp$.
\end{itemize}

We will start by preparing the input matrix $\SG$ for the SVD.  We denote the sine-Gaussians  by the vectors $\s_k$, $k = 1, 2, ...M$; for example, for the parameters considered here and for individual component masses of $7 \Msun$, we have $M$ = 1288.  
Let a data segment of length $T$ be sampled uniformly with $N$ number of points. We find it convenient to work in the Fourier domain. 
Taking the discrete Fourier transform, the samples $\ts_k (f_n)$ in the frequency domain are at the frequencies $f_n = n/T$, where $n$ takes values between $-N/2 \leq n \leq N/2 - 1$. The frequency domain samples $\ts_k (f_n)$ are also $N$ in number and placed $\Delta f = 1/T$ apart in the Fourier space. 
Note each $\s_k \in \D$. Thus $\D$ is $N$ dimensional where $N$ is a large number; we have taken $N = 64 \times 2048 = 131072$ time points -- i.e., points in a data segment of 64 sec. sampled at 2048 Hz.
Thus, $\D$ is practically infinite dimensional (see \citep{DGGB2017} for discussion on this point). We can therefore form a matrix $\SG \equiv G_{kn}$ with rows labelled by $k$ and the columns labelled by $n$; $\SG$ is then a $M \times N$ matrix. The row vectors of $\SG$ are the sine-Gaussians, each having $N$ components in the frequency domain. The matrix $\SG$ has the following form:
\begin{equation}
 \SG = \left[ \begin{array}{cccc} \ts_1 (f_{-N/2}) & \ts_1 (f_{(-N/2 +1)})  & \hdots & \ts_1 (f_{N/2 - 1}) \\ 
 \ts_2 (f_{-N/2}) & \ts_2 (f_{(-N/2 +1)})  & \hdots & \ts_2 (f_{N/2 - 1}) \\
 \vdots & \vdots & \vdots & \vdots \\ 
 \ts_M (f_{-N/2}) & \ts_M (f_{(-N/2 +1)})  & \hdots & \ts_M (f_{N/2 - 1}) \end{array} \right]\,.
 \label{sg_mat}
\end{equation} 
But this is not the matrix that must be used. We need to time-shift each row-vector, namely, the sine-Gaussian $\s_k$, by $-t_d$ and also subtract out the components of the sine-Gaussians parallel to the relevant template $\h$. In order to take care of arbitrary initial phase, we subtract components parallel to both $\h_0$ and $\h_{\pi/2}$. Assuming that the trigger occurs at time zero, we take the match with the templates denoted by $\h_0 (0)$ and $\h_{\pi/2} (0)$. The glitch then must have occurred at time $-t_d$. Then the orthogonal component of the glitch is given by:
\be
\sp (-t_d) = \s (-t_d) - (\s (-t_d), \h_0 (0)) \h_0 (0) - (\s (-t_d), \h_{\pi/2} (0)) \h_{\pi/2} (0) \,. 
\ee
The sine-Gaussian at time $-t_d$ is obtained by multiplying the expression for the sine-Gaussian in the Fourier domain by $e^{2 \pi i f t_d}$. Here $\sp (-t_d)$ denotes the orthogonal component of the time-shifted sine-Gaussian. We have also left out the index $k$ from the row vector in order to avoid clutter. The scalar product on $\D$ (Eq. (\ref{scalar})) has been used. Recall that $t_d$ is a function of $f_0$ and $Q$. Since each row vector in the matrix $\SG$ indexed by $k$   corresponds to a different point in the $(f_0, Q)$ space, each row vector is time-shifted by a different amount. Also the operations of time shifting and taking the orthogonal component can be independently carried out without one affecting the other. This can be easily verified by an explicit computation. From a deeper perspective, the time translation operation can be looked upon as a coordinate transformation. Then the operation of subtracting the parallel component of the glitch is coordinate independent, since it essentially involves a scalar product (the projection) which is invariant under coordinate transformations. We can thus form a matrix with row vectors $\sp_k$ which are both time-shifted and orthogonal to the trigger template. In order that the SVD gives equal weightage to the sine-Gaussians we perform one more operation of normalising the $\sp_k$ so that $\| \sp_k \| = 1$.  We construct the matrix $\SG_\perp$ whose row vectors are $\sp_k, ~k = 1, 2, ..., M$. The vector space spanned by the row vectors of $\SG_\perp$ is precisely $\Vp$ which we have defined above. Note that $\Vp$ is a subspace of $\D$. 
\par
We are not quite done yet. We still need to take cognisance of the scalar product in Eq. (\ref{scalar}) in order that the SVD yields the desired result, 
because the usual SVD algorithm \citep{NumericalRecipes} \citep{Matrix_comp-Frob} assumes an Euclidean scalar product. 
We will take the necessary steps in the next subsection where we will describe how the SVD works and obtain the best lower dimensional approximation to $\Vp$ by invoking the Eckart-Young-Mirsky theorem.

\subsection{Finding the best-fit low-dimensional approximation to $\Vp$}
\label{SVD}

We could in principle use $\Vp$ on which to project the data vector and compute the $\chi^2$ statistic. 
But in practice it would involve too much computational effort and slow down the search pipeline -- 
the $\chi^2$ would involve too many degrees of freedom, namely, the dimension of $\Vp$. In the case of individual masses of $7 \Msun$, the number of d.o.f.s would be over 1000.
We prefer a $\chi^2$ with less than 100 degrees of freedom. 
In order to do this in the best possible manner, we need to compute the best $p$-dimensional approximation to $\Vp$, where $p$ is reasonably small. 
The SVD algorithm allows us to achieve just this -- this is the essence of the Eckart-Young-Mirsky theorem \citep{EY1936}.
 
Consider a set of $M$ vectors in an $N$-dimensional space. In order to seek out an optimal subspace of dimension $p < M$, we have to find a subspace that minimizes the sum of the squares of the perpendicular distances of these $M$ vectors to itself. This is also known as \emph{best least-square-fit} problem. This problem is equivalent to maximizing the sum of the squares of the lengths of projections onto the subspace. We use the \emph{greedy} approach to find the best-fit $p$ dimensional subspace to $\Vp$. Let $\sp_k'$ be the projection of $\sp_k$ onto this $p$-dimensional subspace. Then, we desire a $p$-dimensional subspace of $\Vp$ such that
\begin{align}
\sum_{k = 1}^M \|\sp_k'\|^2    
\end{align}
is maximum -- i.e., the sum of the squares of the projections of $\sp_k$ onto the $p$-dimensional subspace is maximum. The norm used here pertains to the scalar product defined in Eq. (\ref{scalar}). Then this is the subspace $\S$ we  are seeking. We now briefly describe how the SVD works.
 \par
Consider a matrix $\Ab$ of size $M \times N$, where the rows of $\Ab$ are $M$ vectors in an $N$-dimensional space. We define the \emph{first singular vector}, $\textbf{v}_1$, of $\Ab$ as the one that satisfies
\begin{align}
    \textbf{v}_1 = \arg \max_{|\textbf{v}|=1} |\Ab \textbf{v}|\,,
\label{fsnglr}    
\end{align}
where $\v_1$ is an $N$-dimensional column vector in the above equation. Thus, the vector $\textbf{v}_1$ lies along the best-fit line that maximizes $|\Ab \textbf{v}|^2$. We use the modulus notation to signify the Euclidean norm, which is assumed by the usual SVD algorithm. The first (and the largest) singular value is $\sigma_1=|\Ab \textbf{v}_1|$. Now the greedy approach is to take the $\textbf{v}_1$ as the first basis vector and then try to find a unit vector that will maximize $|\Ab \textbf{v}|$ amongst all the vectors perpendicular to $\textbf{v}_1$. Thus, the second singular vector is
\begin{align}
    \textbf{v}_2 = \ \arg \max_{\textbf{v}\perp\textbf{v}_1,|\textbf{v}|=1}|\Ab \textbf{v}|.
\end{align}
Clearly, by definition $|\v_1| = |\v_2 | = 1$ and moreover $\v_1 \cdot \v_2 = 0$. More importantly for our purpose, the 2-dimensional subspace spanned by $\v_1$ and $\v_2$ is the best-fit subspace to the $M$ row vectors constituting the matrix $\Ab$; the vectors $\v_1$ and $\v_2$ form an orthonormal basis of this 2-dimensional subspace. The second singular value is given by $\sigma_2 = |\Ab \v_2 |$.  We can continue in this similar fashion to find the subsequent singular vectors. 
It can be shown that the process eventually stops when one has found 
the $r$ singular vectors $\textbf{v}_1,\textbf{v}_2,...,\textbf{v}_r$. 
We now state the Eckart-Young-Mirsky theorem (without proof)~\citep{EY1936}:

{\bf Theorem}: \emph{Let $\Ab$ be a $M \times N$ matrix where  $\textbf{v}_1,\textbf{v}_2,....,\textbf{v}_r$ are the singular vectors as  defined above. For $1\leq k \leq r$, let $V_k$ be the subspace spanned by $\textbf{v}_1,\textbf{v}_2,....\textbf{v}_k$. Then for each $k$, $V_k$ is the best-fit $k-$dimensional subspace to the vector space spanned by the row vectors of $\Ab$.}

Therefore, the first $k$ singular vectors span the best-fit $k$-dimensional subspace of $\Ab$. The input matrix for the SVD will be taken to be essentially the matrix $\SG_\perp$ but modified in a suitable way in order to account for the {\em weighted} scalar product. The SVD decomposition of $\Ab$ is written in the form:
\begin{align}
\Ab = \Ub ~\Sgm ~ \Vb^{\dagger}  \,,  
\end{align}
where $\Ab$ is an $M \times N$ matrix, $\Ub$ is the $M \times r$ matrix of left singular vectors, $\Sgm$ is an $r \times r$ square diagonal matrix of singular values $\sigma_1, \sigma_2, ..., \sigma_r$ arranged in descending order of magnitude and $\Vb^{\dagger}$ is the $r \times N$ matrix of right singular vectors. The superscript dagger on $\Vb$ denotes the Hermitian conjugate of $\Vb$. The left and right singular vectors are normalised and are arranged as column vectors in the matrices $\Ub$ and $\Vb$, respectively. Our main interest lies in the matrices $\Vb$ and $\Sgm$, which we will judiciously truncate to obtain the best-fit subspace $\S$ to the desired level, based on the singular values $\sigma_k$. 
\par
The question is at what $k$ do we truncate? For this purpose we define the Frobenius norm \citep{Matrix_comp-Frob} of the matrix $\Ab$ to be:
\begin{align}
\| \Ab \|_F^2 = \sum_{i = 1}^M \sum_{j = 1}^N |a_{ij}|^2 \equiv \sum_{k = 1}^r \sigma_k^2 \,.
\end{align}
The Frobenius norm of a matrix $\Ab$, denoted by the subscript $F$ gives the full content of the matrix which is also summed up in terms of its singular values $\sigma_k$. Suppose we decide on 90$\%$ level of accuracy, then we choose $p$ so that $\sum_{k = 1}^p \sigma_k^2 ~\gtrsim ~0.9 ~\| \Ab \|_F^2$. We define $\S$ as the span of the first $p$ right singular vectors $\v_1, \v_2, ..., \v_p$; in fact they constitute an orthonormal basis of $\S$. This also means that the sum of squares of projections of the row vectors of $\Ab$ on $\S$ add up to more than 90$\%$ of the full value. If a glitch vector is close to any of these row vectors, its square of the norm of its projection onto $\S$ will tend to be large, which will result in a large $\chi^2$. This is in fact the goal we started with.  
\par
We now turn to the final aspect of how the weighted scalar product can be included into the SVD machinery so that it gives the desired results. 
We first give the prescription and then justify it. We start with the matrix $\SG_\perp$.
We go to the frequency domain and divide each entry of $\SG_\perp$ corresponding to a frequency $f_k$ by $\sqrt{S_h (|f_k|)}$. 
We have taken the modulus because the frequency ranges from negative to positive values. Recall that we are dealing with a one-sided PSD, which therefore obeys $S_h (-f_k) = S_h (f_k)$.  Accordingly, we construct the ``whitened" sine-Gaussian matrix $\SG_W$ as follows:
\begin{equation}
 \SG_W = \left[ \begin{array}{ccc} \frac{\ts_{\perp 1} (f_{-N/2})}{\sqrt{S_h (|f_{-N/2}|)}}  & \hdots & \frac{\ts_{\perp 1} (f_{N/2 - 1})}{\sqrt{S_h (f_{N/2 - 1})}} \\ 
 \vdots & \vdots & \vdots \\ 
 \frac{\ts_{\perp M} (f_{-N/2})}{\sqrt{S_h (|f_{-N/2}|)}} & \hdots & \frac{\ts_{\perp M} (f_{N/2 - 1})}{\sqrt{S_h (f_{N/2 - 1})}} \end{array} \right]\,.
 \label{sgw_mat}
\end{equation} 
Next we perform the SVD of $\SG_W$ and write:
\begin{align}
\SG_W = \Ub_W ~\Sgm_W ~ \Vb_W^{\dagger}  \,,  
\end{align}
where the subscript $W$ denotes the corresponding whitened matrices. We now consider $\Vb_W^{\dagger}$ and unwhiten its rows. Denoting the entries of $\Vb_W^{\dagger}$ by $v'_{ij}$, where the index $i$ runs over the frequency index from $-N/2$ to $N/2 - 1$ and $j = 1, 2, ..., r$, we get the unwhitened matrix $\Vb^{\dagger}$ by setting $v_{ij} = v'_{ij} \sqrt{S_h (f_i)}$. The right singular vectors are the columns of $\Vb$. We just choose the first $p$ of these singular vectors so that they give the desired level of accuracy. Then these $p$ vectors form an orthonormal basis of $\S$ and generate $\S$.
\par
It now remains to justify our whitening procedure that we have used above, when we have a general scalar product as defined in Eq. (\ref{scalar}). The scalar product of Eq. (\ref{scalar}) can be written in the form:
\begin{align}
(\v, \w) = \sum_{k = 1}^N \mu_k v_k^* w_k \,,
\label{scalar_2}
\end{align}
where $\mu_k > 0$ are positive real numbers. We may regard the vectors being decomposed in a Fourier basis with $\mu_k = (S_h (f_k))^{-1}$. Then essentially the scalar product defined by Eq. (\ref{scalar_2}) is a discretised version of Eq. (\ref{scalar}). (Note that here for convenience, we have labelled the Fourier components from $1$ to $N$ instead of $-N/2$ to $N/2 - 1$. We are free to do this.) The norm of a vector $\v$ is given by $\| \v \|^2 = \Sigma_{k = 1}^N \mu_k |v_k|^2$. Let a $M \times N$ matrix $\Ab$ be given. Let us find the first singular vector $\v$ of $\Ab$ with the scalar product (\ref{scalar_2}). Then the vector $\v$ must be of unit norm and should be such that $\| (\Ab \v) \|^2$ is maximum; that is, we must maximise:
\begin{equation}
\sum_{i = 1}^M \left (\sum_{j = 1}^N \mu_j a_{ij}^* v_j \right )^2 ~~~~ {\rm subject~ to} ~~~~
\sum_{k = 1}^N \mu_k |v_k |^2 = 1. 
\end{equation}
Note that each sum over the index $j$ is real, because each row vector of the matrix $\Ab$ and $\v$ are real vectors, although expressed in a complex Fourier basis. So we have a sum of $M$ terms which are squares of real numbers and this sum needs to be maximised. But this problem can be readily mapped to that of the Euclidean scalar product by defining a matrix $\Ab'$ with entries $a'_{ij} = \sqrt{\mu_j} a_{ij}$ and  vectors $\v'$ by $v'_i = \sqrt{\mu_i} v_i$. In the primed variables we need to maximise $|\Ab'\cdot \v'|^2$ subject to $\v'\cdot \v' = 1$, where the ``dot" represents the usual Euclidean scalar product. The first singular vector is then $\v'_1$ and from it we can obtain the corresponding $\v_1$ for the original problem by writing $v_i = v'_i /\sqrt{\mu_i}$. This procedure can be continued to obtain the subsequent singular vectors in a similar way. This argument justifies the whitening procedure we have adopted above. 

\section{Results}
\label{sec:results}

We next apply the paradigm developed above to test if the
optimal sine-Gaussian $\chi^2$ statistic
actually provides any additional power in distinguishing CBC signals from transient noise artifacts. 
To be able to interpret the results, we continue to model the artifacts as sine-Gaussians, with various values for their quality factor and central frequency. We apply them to real data with real glitches in a subsequent work~\cite{Choudhary2020}.
We use the PyCBC Software~\citep{pycbc-software,Allen:2005fk,Usman:2015kfa,Nitz:2017svb} for searching for simulated BBH signals.

\subsection{The parameter space of signals and glitches}

All of our CBC signal and noise artifact injections are made in simulated Gaussian data with aLIGO ZDHP \citep{alzdhp_psd} as the noise PSD and a lower frequency-cutoff of 20 Hz.
In the realm of signals, we limit ourselves to injections of simulated non-spinning CBC signals -- all modeled with the IMRPhenomP waveform approximant~\citep{Hannam:2013oca,imrphenompv3} -- with component masses $m_{1,2}\in [7,100]M_\odot$ and total mass $M \equiv (m_1 + m_2) \in [14,160]M_\odot$.
To search for signals in these simulations we employ two kinds of template banks: (a) The Full bank: This bank has templates that cover the parameter space of the CBC injections we chose for our study fully. (b) The Targeted banks: Parallelly, we search for the same signals with multiple small banks, each of which covers a subset of the full $m_{1,2}$ space. These are called targeted banks. They are designed so that they overlap with each other in the $m_{1,2}$ space so as not to lose signals with parameter values at the boundaries of each of those banks. For both kinds of banks we require a minimal match of 97\% among neighboring templates with a lower frequency-cutoff of 20 Hz.
The parameter ranges of these template banks are listed in Table  \ref{tab:Template-bank_parameters}.

\begin{figure}[H]
        \centering
       \includegraphics[scale = 0.95]{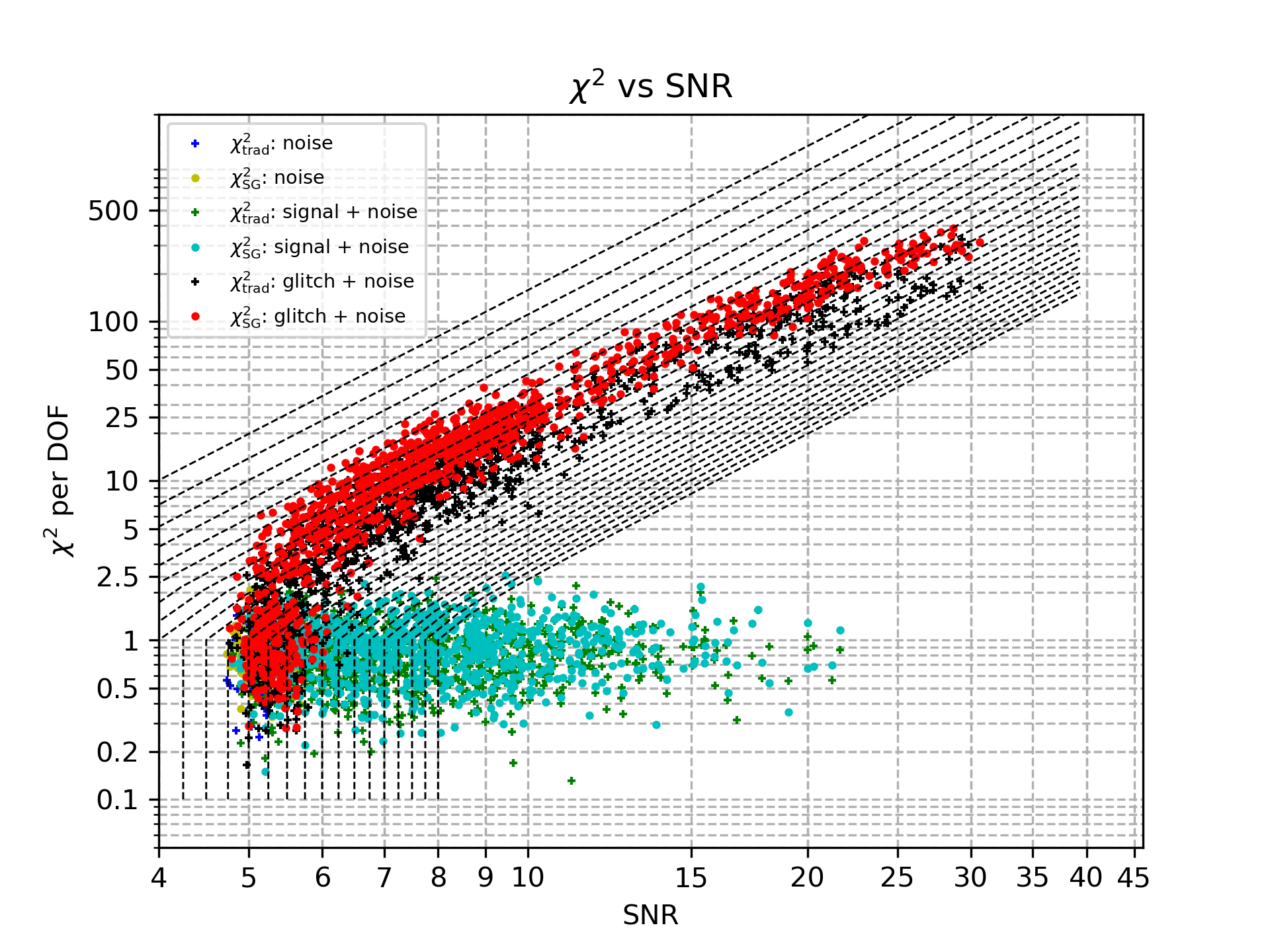}

    \caption{The traditional and optimal ``SG" $\chi^2$ statistics (see legend), per degree of freedom, are plotted {\it vs} SNR for various types of triggers. 
    These arise from injections of simulated (a) noise (Gaussian), (b) glitches (sine-Gaussians, of the high-$Q$-low-$f_0$ type defined in Table~\ref{tab:Glitch_parspace})
    and (c) BBH signals, of the category 6 type defined in Table~\ref{tab:Injection_mass_ranges}, when employing Targeted template bank 3, as described in Table~\ref{tab:Template-bank_parameters}. The ROC curves for these triggers are shown in the left plot in Fig.~\ref{fig:roc-highQlowFm1}.}
    \label{fig:ch2vsnr-highQlowFm1}
\end{figure}

For assessing the effect of noise artifacts, and even plain Gaussian noise (with aLIGO ZDHP noise PSD), in our searches, we match-filter simulated data with these features against the same template banks and compute both the SNRs and the $\chi^2$ -- both the traditional $\chi^2$ and our sine-Gaussian $\chi^2$. 
These are shown for various cases in Fig. \ref{fig:ch2vsnr-highQlowFm1}. As expected, these plots show that CBC triggers and noise triggers separate cleanly for large SNRs but not for small SNRs, which is expected. This is true regardless of the type of $\chi^2$ employed. This aspect notwithstanding, some figures present qualitative evidence for the fact that the noise artifacts register higher sine-Gaussian-$\chi^2$ values than traditional $\chi^2$ values, even if slightly. This in itself is not proof that the former $\chi^2$ is a better discriminator here. To establish that possibility one needs to assess what the $\chi^2$ values are (for both kinds of statistics) for the CBC signals as well as the noise artifacts. This comparison is best done, quantitatively, with Receiver-Operating Characteristic (ROC) curves, which we discuss below.


\begin{figure}[H]
        \centering
       
       \includegraphics[scale = 0.5]{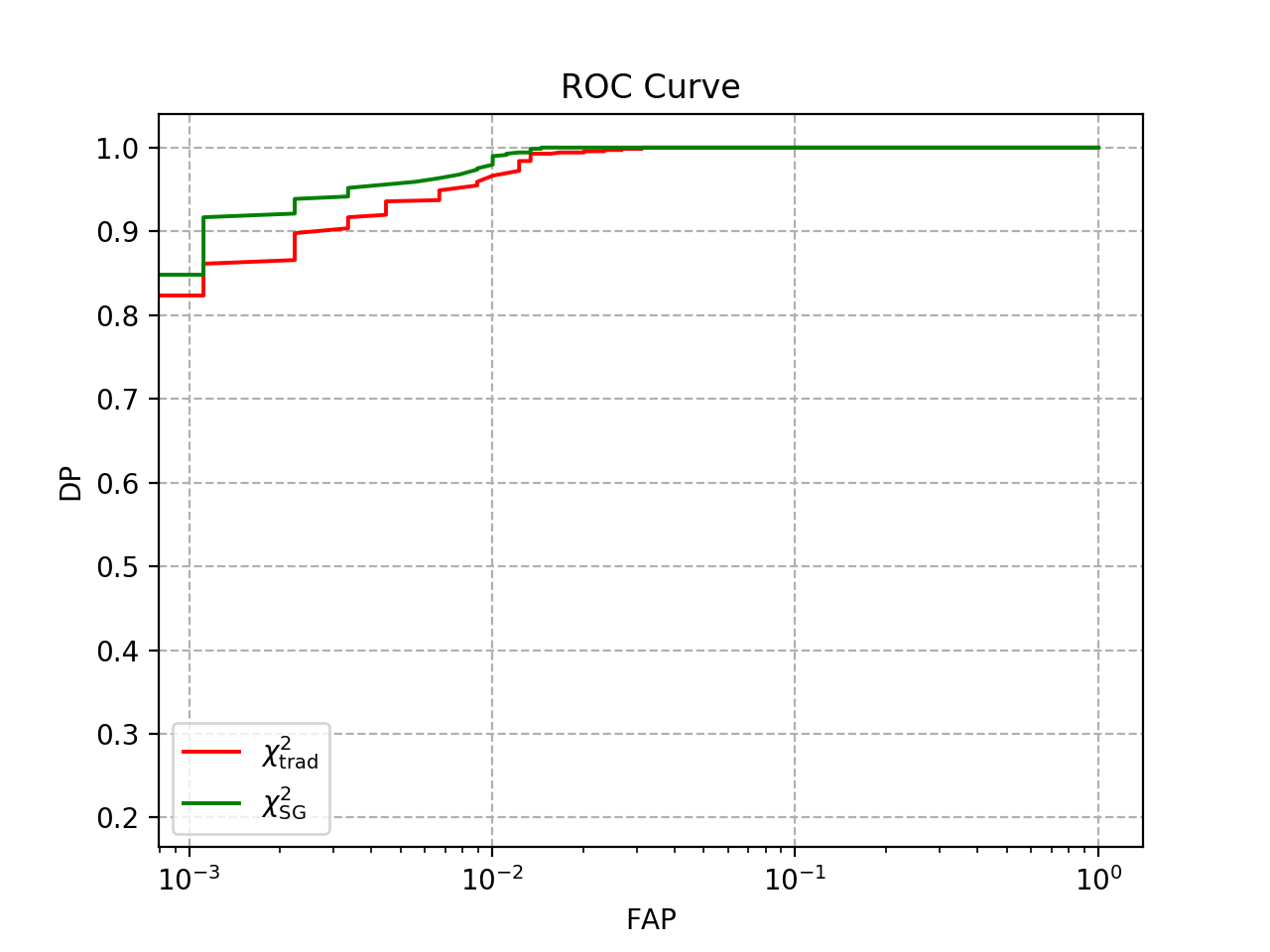}
       \includegraphics[scale = 0.5]{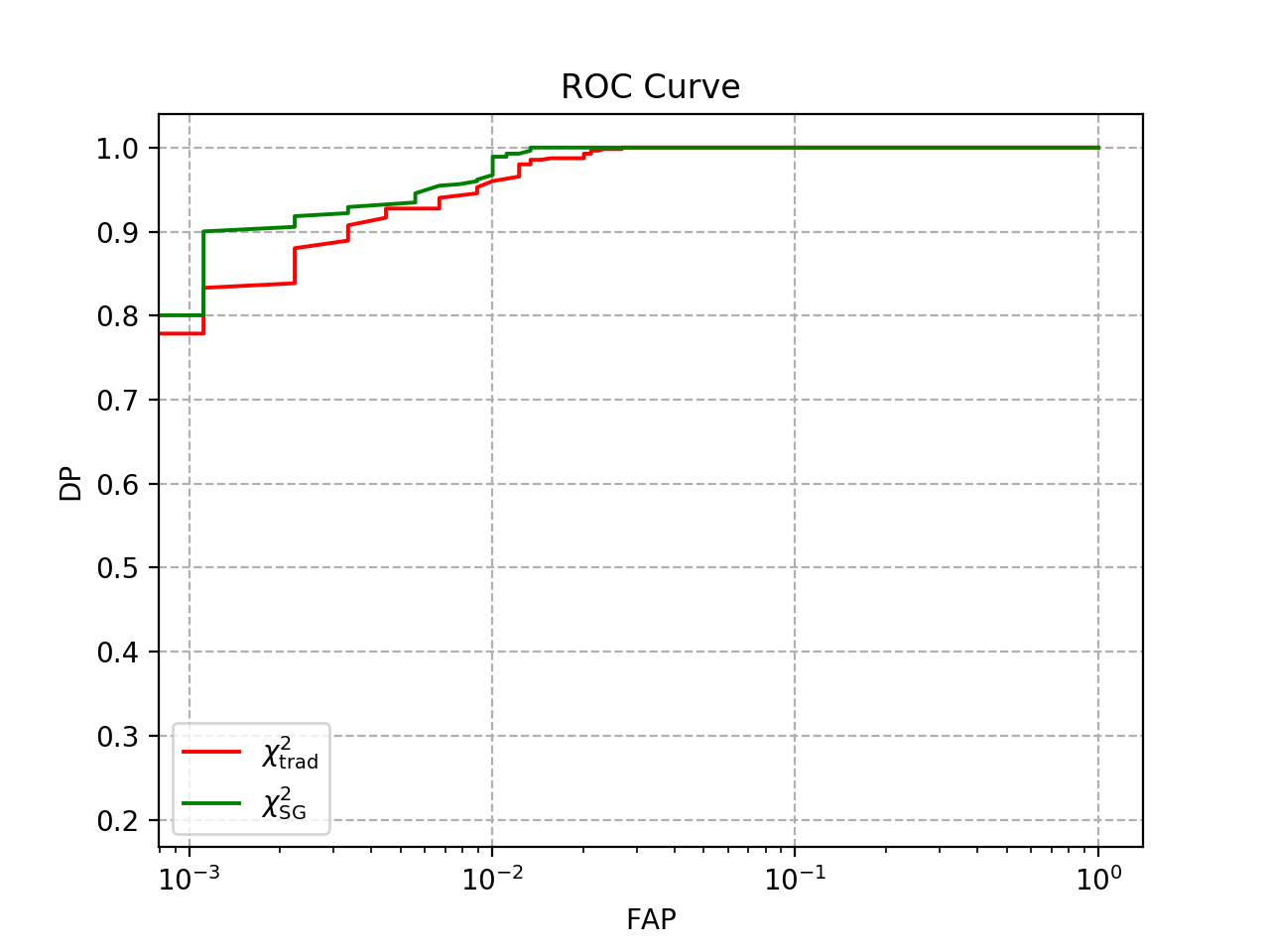}

    \caption{In the left plot, we show the ROC curves comparing performances of the same two $\chi^2$ statistics and triggers as in Fig.~\ref{fig:ch2vsnr-highQlowFm1}.
    The right plot is a similar comparison, for the same Gaussian noise and sine-Gaussian glitch triggers
    but for BBH injections of category 5 of Table~\ref{tab:Injection_mass_ranges}, using Targeted bank 3 of Table~\ref{tab:Template-bank_parameters}.}
    \label{fig:roc-highQlowFm1}
\end{figure}

\begin{figure}[H]
        \centering
        \includegraphics[scale = 0.5]{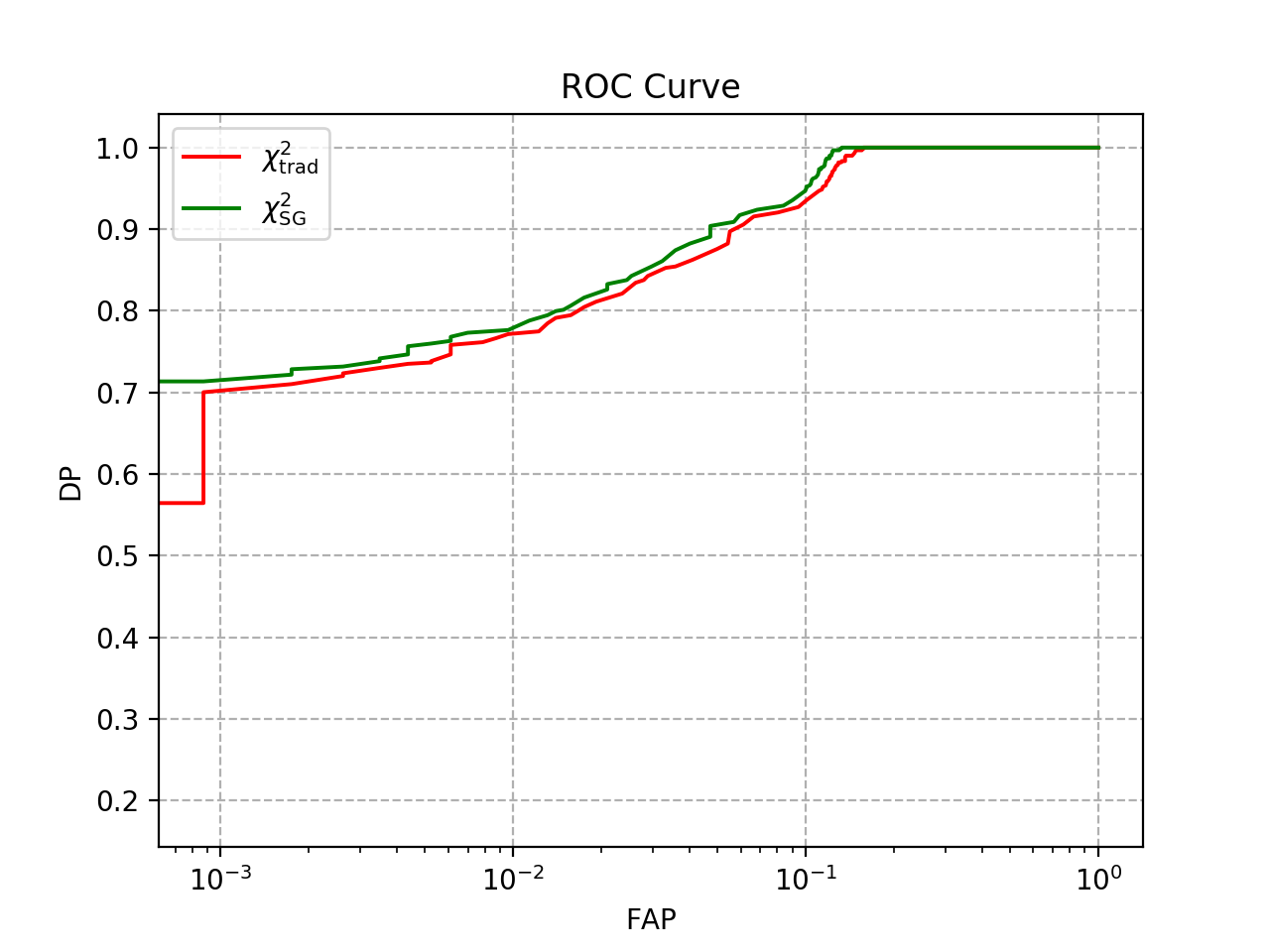}
        \includegraphics[scale = 0.5]{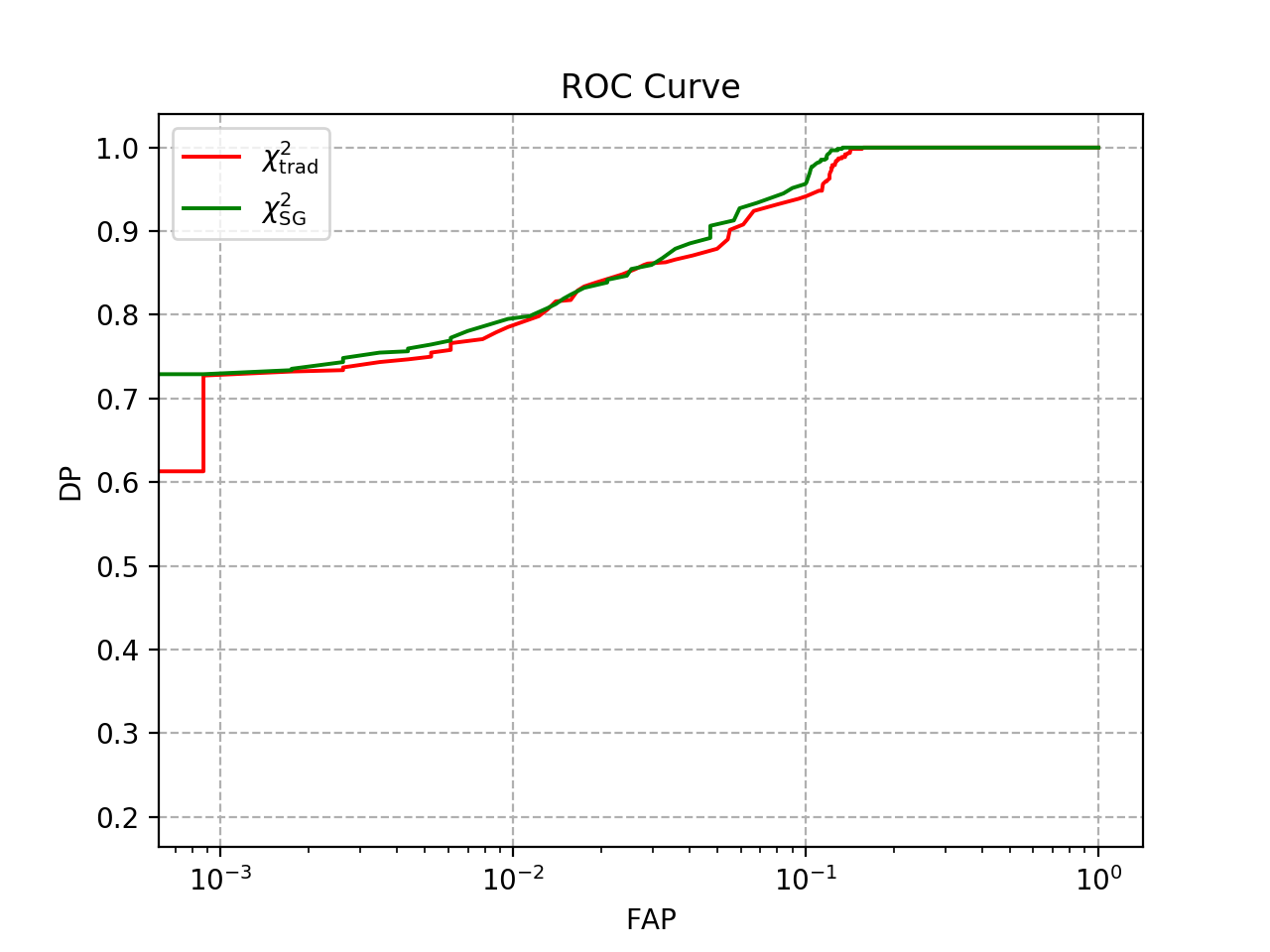}

    \caption{ROC curves comparing performances of the two $\chi^2$ statistics and triggers
    as in Fig.~\ref{fig:ch2vsnr-highQlowFm1}, except 
    for the BBH injections, which are of categories 3 (left) and 4 (right) of Table~\ref{tab:Injection_mass_ranges}, both using Targeted bank 2 of Table~\ref{tab:Template-bank_parameters}.
    The sudden drop in the ROC for $\chi^2_{\rm trad}$ around a FAP of $10^{-3}$ most likely arises owing to the difficulty of producing enough loud triggers purely from noise, and should be interpreted with care when comparing with the other ROC curve.
    }
    \label{fig:roc-highQlowFm2}
\end{figure}

In order to construct an ROC curve, we first define a new detection statistic that is derived from the SNR ($\rho$) and $\chi^2$ as follows: 
\begin{eqnarray}
    \rho_{\rm OSG} &=& \rho \,, \quad\quad\quad\quad\quad\quad \quad\quad\quad\quad\quad\quad\>
    \chi^2_r \leq 1,
    \label{rhoOSG1}\\
    &=& \rho \left[ \frac{1}{2} \left(1 + \left(\chi^2_r \right)^3\right) \right]^{-1/9}\,, 
    \quad\quad\quad
    \chi^2_r > 1,
    \label{rhoOSG2}
\end{eqnarray}
where $\chi^2_r$ is just the $\chi^2$ per degree of freedom, for both the traditional and optimal sine-Gaussian kind.
The new statistic above resembles the re-weighted SNR~\cite{Babak:2005kv,Nitz:2017lco}, except that in the latter, the exponent of $-1/9$ in Eq.~(\ref{rhoOSG2}) is replaced by $-1/6$.
The detection probability (DP) at any given value of $\rho_{\rm OSG}$ is the fraction of all triggers associated with simulated BBH signals that are found with a new detection statistic value that is larger. On the other hand, the 
False-Alarm Probability (FAP) corresponding to that $\rho_{\rm OSG}$ value is the fraction of triggers from noise or glitches that have a new detection statistic value greater that that reference.
The contours of the so computed constant FAP of 
$\rho_{\rm OSG}$ 
are overlaid with  dashed black lines in the $\chi^2_r$ {\em vs} SNR plot in Fig.~\ref{fig:ch2vsnr-highQlowFm1}.
The plot of DP {\em vs} FAP for any detection statistic is its ROC curve. Such curves for $\rho_{\rm OSG}$ (with the optimal sine-Gaussian $\chi^2_r$) and the re-weighted SNR (with the traditional $\chi^2_r$) are compared for various categories of simulations in Figs.~\ref{fig:roc-highQlowFm1} and \ref{fig:roc-highQlowFm2}.

The main results brought forth by the ROC curves are as follows. In essentially all cases, the performance of optimal $\chi^2$ in recovering CBC signals at any SNR (or FAP) studied is comparable to or better than that of the traditional $\chi^2$, even if by a small degree. The improvement is often by a few to several percent, especially, near a FAP of $10^{-3}$.
Alternatively, at the same detection probability the false-alarm probability of a BBH signal is perceptively lower for the new $\chi^2$ statistic.
Recall that for the traditional $\chi^2$, the detection statistic used in these comparisons was the re-weighted SNR,
as is customary. If we use it with $\rho_{\rm OSG}$, then the optimal $\chi^2$ performs much better than the traditional one, sometimes by 10 - 15 \% (not shown), near a FAP of $10^{-3}$. With better tuning, the performance of the new $\chi^2$ may show further improvement. We plan to pursue this in real data.


\begin{table}
    \centering
    \begin{tabular}{|C{3cm}|C{2cm}|C{2cm}|C{2cm}|C{2cm}|C{2cm}|}
        \hline
         \textbf{Bank} & $\bf m_{min} (M_{\odot})$ & $\bf m_{max} (M_{\odot})$ & $\bf M_{min} (M_{\odot})$ & $\bf M_{max} (M_{\odot})$ & \textbf{No. of templates}\\
         \hline
         Full bank & $5$ & $100$ & $10$ & $165$ & $2222$\\
         \hline
         Targeted bank 1 & $5$ & $40$ & $10$ & $45$ & $1603$\\
         \hline
         Targeted bank 2 & $5$ & $78$ & $35$ & $85$ & $845$\\
         \hline
         Targeted bank 3 & $5$ & $100$ & $75$ & $125$ & $297$\\
         \hline
         Targeted bank 4 & $5$ & $100$ & $115$ & $165$ & $16$\\
        \hline
    \end{tabular}
    \caption{Template-bank parameters: The ranges of various source parameters that characterize the template banks used in our studies. Above, $m_{\rm min}$, $m_{\rm max}$ are the lower and upper bounds on the component masses of the binary, respectively. On the other hand, $M_{\rm min}$ and $M_{\rm max}$ are the lower and upper bounds on the the total mass of the binary, respectively. }
    \label{tab:Template-bank_parameters}
\end{table}


\begin{table}
    \centering
    \begin{tabular}{|C{3cm}|C{2cm}|C{2cm}|C{2cm}|C{2cm}|}
         \hline
          & $\bf Q_{min}$ & $\bf Q_{max}$ & $\bf f_{0~min} (Hz)$ & $\bf f_{0~max} (Hz)$ \\
         \hline
         high $Q$, low $f_0$ & $25$ & $50$ & $40$ & $80$ \\
         \hline
         high $Q$, high $f_0$ & $25$ & $50$ & $80$ & $120$ \\
         \hline
         low $Q$, low $f_0$ & $5$ & $15$ & $40$ & $80$ \\
         \hline
         low $Q$, high $f_0$ & $5$ & $15$ & $80$ & $120$ \\
         \hline
    \end{tabular}
    \caption{The glitch injections used in our study are all sine-Gaussians, and were grouped into the four categories described in the rows above. They are all parameterized by $Q$ and $f_0$, with varying ranges as tabulated here.  }
    \label{tab:Glitch_parspace}
\end{table}



\begin{table}[H]
    \centering
    \begin{tabular}{|c|c|c|c|c|c|}
        \hline
         Sr. no. &  $\bf m_{min} (M_{\odot})$ & $\bf m_{max} (M_{\odot})$ & $\bf M_{min} (M_{\odot})$ & $\bf M_{max} (M_{\odot})$ & Average $p$\\
        \hline
         1 & $7$ & $21$ & $14$ & $28$ & 28\\
        \hline
         2 & $7$ & $35$ & $28$ & $42$ & 18\\
        \hline
         3 & $7$ & $53$ & $42$ & $60$ & 15\\
        \hline
         4 & $7$ & $73$ & $60$ & $80$ & 14\\
        \hline
         5 & $7$ & $93$ & $80$ & $100$ & 13\\
        \hline
         6 & $7$ & $100$ & $100$ & $120$ & 12\\
        \hline
         7 & $7$ & $100$ & $120$ & $140$ & 13\\
        \hline
         8 & $7$ & $100$ & $140$ & $160$ & 13\\
        \hline
    \end{tabular}
    \caption{Parameters of the simulated signals used in our injection studies are divided into above ranges. Above, $m_{\rm min}$, $m_{\rm max}$ are the lower and upper bounds on the component masses of the binary, respectively. On the other hand, $M_{\rm min}$ and $M_{\rm max}$ are the lower and upper bounds on the the total mass of the binary, respectively. $p$ is the dimension of the orthogonal subspace on which the $\chi^2$ is defined.}
    \label{tab:Injection_mass_ranges}
\end{table}




\section{Conclusions}
\label{concl}

\par

In this work we have constructed a $\chi^2$ statistic that is optimally effective in discriminating BBH signals from  sine-Gaussian glitches and, more broadly,  glitches that have strong overlap with sine-Gaussians. Past authors have devised signal-based $\chi^2$ discriminators that have been quite successful in identifying triggers arising from noise artifacts in the data (see, e.g., Refs.~\cite{Allen2004,Babak:2005kv,HannaThesis2008,Bose:2011km,Harry:2010fr,Talukder:2013ioa,DGGB2017,Nitz:2017lco,Dupree:2019jqn} and the references therein).
Lately however, their weaknesses, especially in high-mass BBH searches, has become more evident. 
This realization has led to new proposals for reducing their impact on BBH search sensitivities.

Reference~\cite{DGGB2017} for the first time developed the proper mathematical formalism for geometrically understanding existing
signal-based
$\chi^2$ discriminators and constructing new ones. It also showed how one can naturally and unambiguously combine multiple signal-based
$\chi^2$s. 
In the context of the current paper, Ref.~\citep{DGGB2017} provided a formalism for exploiting the characteristics of noise artifacts to construct $\chi^2$ discriminators targeting them. 
Here we have followed up on this idea and gone further with the construction of the optimal $\chi^2$ for sine-Gaussian glitches. However, we find that there are several involved steps that need to be taken before one arrives at that final goal. We briefly outline those steps below. We first consider a family of sine-Gaussian strain snippets in a given physical range of parameters, which we have called $\G$. We then sampled $\G$ uniformly by using a metric so that it is adequately represented. Care has to be taken to time-delay the sine-Gaussians 
in the sampling process. However, it turns out that the number of sampled glitch vectors for $\G$ is too large and consequently the subspace $\VG$ spanned by them also has high dimensionality. 
A low-dimensional approximation to $\VG$ is sought in order that the computational costs for the $\chi^2$ remain in control. The best possible low-dimensional approximation to $\VG$ is obtained by invoking the Eckart-Young-Mirsky theorem and is achieved with the help of the SVD algorithm. Further we ensure that the associated subspace obtained for the $\chi^2$ is orthogonal to the trigger template by appropriately projecting out the components of the glitch vectors
parallel to the trigger template. Carrying out the above steps results in the required optimal $\chi^2$ discriminator for sine-Gaussians - the  $\chi^2_{\rm SG}$. 
We remark that this procedure may seem computationally expensive since $\S_{\rm SG}$ is required at each template in the bank. However, it may be noted that, $\S_{\rm SG}$ at any given template is needed only approximately. We may therefore envisage an interpolation scheme by which $\S_{\rm SG}$ is precomputed only on a coarse grid of the parameter space and it is obtained for any intermediate template by interpolation techniques.

A recent paper~\cite{Nitz:2017lco} proposes a somewhat different way of constructing a $\chi^2$ discriminator that targets a specific type of glitch -- namely ``blips"~\cite{Cabero:2019orq}. Blip glitches are found to have significant projections on a certain subset of sine-Gaussians. 
A set of 20 sine-Gaussian basis vectors -- all with $Q=20$ -- was used 
to construct that subspace.
In this alternative method one constructs a $\chi^2$-like statistic without  subtracting the BBH template or orthogonalizing the sine-Gaussian basis vectors. For that reason, strictly speaking, such a statistic does not have a $\chi^2$ distribution. Moreover, it cannot be unambiguously combined with other $\chi^2$ statistics to improve search sensitivity. The $\chi^2$ statistic proposed here does not suffer from those problems and can be readily implemented in real data.

As mentioned above, in an upcoming work~\cite{Choudhary2020} that implements our optimal $\chi^2$ statistic in real data, we will compare its performance on blip glitches as well. It is conceivable that our statistic may need to be tuned to optimize its performance on this particular kind of glitch, e.g., by specifying how to select the subset of sine-Gaussian basis vectors. Note, however, that our statistic is more general in its applicability than just blips. It should also work on other glitches that have good projections on sine-Gaussians. We plan to test this prospect as well in real data. Here we have taken the first steps toward realizing that goal by  
illustrating the implementation of our $\chi^2_{\rm SG}$ on simulated glitches, BBH signals and Gaussian detector noise (with aLIGO-ZDHP PSD). Through the construction of $\chi^2$ {\it vs} SNR plots and ROC curve comparison we find that 
incorporating the $\chi^2_{\rm SG}$ statistic in BBH searches improves detection probability for several mass ranges compared to the traditional $\chi^2$.
The improvement is  manifest for BBH signals, for various masses -- listed in Table~\ref{tab:Injection_mass_ranges} -- and is by a few to several percentage points. 
That table also shows how the dimensionality of the sine-Gaussian subspace utilized for the optimal sine-Gaussian $\chi^2$ construction varies with the template masses. 
Note that this dimensionality is not very large, which makes its implementation computationally viable. This study prepares us to make the case for utilizing prioritized computing resources for deploying this search statistic in real data. 

It may be observed that there is no dramatic increase in the value of the $\chi^2$ from the traditional to the optimal sine-Gaussian. This is because we have focussed on a particular type of glitch, namely, the  sine-Gaussian glitch, which is ubiquitous. 
Our selection of the sine-Gaussian glitch was motivated from this physical reason. Our results, in fact, show that the traditional $\chi^2$ does pretty well on these types of glitches; of course, our sine-Gaussian $\chi^2$ does better, as it should, since it is by construction optimal for this type of glitch. From the mathematical point of view, the glitches have good projection on subspace $\S_{\rm trad}$ associated with the traditional $\chi^2$ and best projection on an average on the sine-Gaussian subspace $\S_{\rm SG}$. 
However, one could conceive of another type of glitch, say glitch $X$, which is orthogonal (or nearly so) to $\S_{\rm trad}$. Then the traditional $\chi^2$ would be small and thus ineffective in ruling out the $X$-glitch. But in the unified $\chi^2$ formalism, one can always construct an optimal $\chi_{X}^2$ with the associated subspace $\S_X$, by carrying out an analogous procedure as was employed here for the sine-Gaussians. 
Such a $\chi^2$ would optimally rule out the $X$ glitches. Our aim was to point out the generality of our constructive procedure which can be applied to a different family of glitches for which the traditional $\chi^2$ was ineffective. Such glitches may well exist in the data or reveal themselves as detectors are commissioned in the future.

We also remark that employing $\chi^2_{\rm SG}$ does not preclude the application of other $\chi^2$s. In fact in Ref.~\cite{DGGB2017} it has been argued that one can sensibly combine several $\chi^2$s just by adding their associated subspaces $\S$ -- in the vector-space sense -- and construct a combined $\chi^2$. The resulting statistic would discriminate against all the glitches for which each $\chi^2$ was designed. For example, we may add the associated subspaces $\S_{\rm trad}$ and $\S_{\rm SG}$ to form the new subspace $(\S_{\rm trad} + \S_{\rm SG})$, which results in a more powerful $\chi^2$ that can discriminate against glitches for which the traditional $\chi^2$ is optimal as well as those for which the sine-Gaussian $\chi^2$ is optimal. Such a combined statistic will be very useful in reducing false alarms and, thereby, improve the overall significance of GW events. 

\section{Acknowledgments}

Prasanna Joshi would like to thank Shomik Adhicary, Raj Patil, Palash Singh and Rahul Poddar for helpful discussions. Rahul Dhurkunde would like to thank Sourath Ghosh, Sunil Choudhary and Sudhagar.S for helpful discussions. Thanks are due to Bhooshan Gadre for carefully reading the manuscript and making helpful comments.
Many of the simulations reported here were carried out at the IUCAA computing cluster Sarathi.  SVD acknowledges the support of the Senior Scientist Platinum Jubilee Fellowship from 
NASI. We thank Tata Trusts for partial funding support of this work. This document has been assigned the preprint number LIGO-P2000194.

\bibliography{refs}   
\end{document}